%% file: GR=RG-Final.tex
\documentclass[aps,prd,amsmath,amssymb,nofootinbib, 12pt]{revtex4-2}

\usepackage{graphicx}
\usepackage{xcolor}
\usepackage{dcolumn}
\usepackage{bm}
\usepackage{titlesec}
\usepackage{amsthm,physics,hyperref,mathtools,dsfont}
\usepackage[capitalise]{cleveref}
\usepackage{bbm}

\usepackage[normalem]{ulem}
\usepackage{tikz}
\usepackage{tikz-feynman}
\usepackage{subcaption}
\usepackage{ulem}
\usepackage{amsmath}
\usepackage{lmodern}
\usepackage[scr=boondox]{mathalpha}
\DeclareMathAlphabet{\mathpzc}{OT1}{pzc}{m}{it}

\tikzfeynmanset{compat=1.1.0}

\input{macro}

\begin{document}

\title{
Cardinal RG=GR:
Phase-Space Cardinality and the\\ Cosmological Constant and Hierarchy Problems  }

\author{Laurent Freidel}\email{lfreidel@perimeterinstitute.ca}
\affiliation{Perimeter Institute for Theoretical Physics, 31 Caroline St. N., Waterloo ON, Canada}
\author{Jerzy Kowalski-Glikman}\email{jerzy.kowalski-glikman@uwr.edu.pl}
\affiliation{National Centre for Nuclear Research, ul. Pasteura 7, 02-093 Warsaw, Poland}
\affiliation{University of Wroc\l{}aw, Faculty of Physics and Astronomy, pl.\ M.\ Borna 9, 50-204
Wroc\l{}aw, Poland}
\author{\fbox{Robert G. Leigh}}
\affiliation{Illinois Center for Advanced Studies of the Universe \& Department of Physics,
			University of Illinois, 1110 West Green St., Urbana IL 61801, U.S.A.}
\author{Djordje Minic}\email{dminic@vt.edu}
\affiliation{Department  of Physics, Virginia Tech, Blacksburg, VA 24061, U.S.A.}

\date{\today}

\begin{abstract}
The
effective field theory approach constructs the low-energy theory by integrating out fluctuations above an ultraviolet scale
which raises the cosmological constant and hierarchy problems.
We investigate an alternative prescription in which the quantum theory is restricted to a finite accessible region of phase space, characterized by a dimensionless cardinality $\mathcal N$, rather than by independently prescribed spacetime and momentum scales. At fixed cardinality these scales are related. For Euclidean de Sitter space, the stationarity condition for the gravitational action can then be written as a stationarity condition with respect to the momentum scale, a relation which we call Cardinal RG=GR. We demonstrate the relation between $\mathcal N$ and the Gibbons-Hawking entropy of de Sitter space, compute the cosmological constant proportional to ${1}/{\sqrt{\mathcal N}}$, show that it is radiatively stable and apply the same prescription to demonstrate the radiative stability of the mass parameter for a simple interacting scalar field model, thereby illuminating the resolution of the hierarchy problem. We also establish a precise limit, involving gravitational data, in which the conventional results of Quantum Field Theory and the Renormalization Group are recovered.

\end{abstract}

\maketitle

\newpage

\newpage

{\bf Dedication}: This paper was completed while our dear friend and long-time collaborator Rob Leigh (1964–2026) was battling a devastating illness that ultimately claimed his life on April 18, 2026. Rob was an extraordinary person and an exceptional physicist, without whom we could not imagine our personal lives or our scientific work. Rob was involved in the original discussions that ultimately led to this paper, but due to his illness he could not participate in its actual development. Nevertheless he was present in spirit every step of the way. We dedicate this paper to his lasting memory and aspire to honor his visionary, broad-thinking and  forward-looking spirit.

\section{Introduction and Overview}\label{Sect:Introduction}
The standard effective field theory (EFT) formulation of the Standard Model~\cite{Weinberg:1968de, Coleman:1969sm, Callan:1969sn, Wilson:1971bg, Wilson:1971dh, Wilson:1973jj, Wilson:1974mb, Weinberg:1978kz, Polchinski:1983gv, Wilson:1983xri, Weinberg:2009bg} (see, for instance,~\cite{Isidori:2023pyp} for a recent review) encounters two celebrated problems: the vacuum energy, or cosmological constant, problem~\cite{Weinberg:1988cp, Weinberg:2000yb} and the hierarchy problem associated with the Higgs boson mass~\cite{Craig:2022eqo, Peskin:2025lsg}, and more generally with scalar masses. In the usual perturbative treatment these problems appear as the familiar quartic and quadratic sensitivity to high energy scales\footnote{In the words of Steven Weinberg: ``These are great mysteries that confront us: Why are the terms in our present theory that have the dimensions of positive powers of mass so small compared to the scale that we think is the fundamental scale, somewhere in the neighborhood of $10^{15}-10^{18}$ GeV? We don’t know.''~\cite{Weinberg:2021exr}.}. They are usually formulated as naturalness problems, namely as a question why the observed parameters remain small in the presence of large quantum corrections~\cite{Padilla:2015aaa}. In this paper we take a somewhat different point of view. We regard both problems as connected and as indications that the usual separation between quantum fields and the spacetime geometry on which they live becomes incomplete when gravity is included.
(This point has been also emphasized, from a different viewpoint, in~\cite{Becker:2020mjl, Becker:2021pwo, Ferrero:2024yvw, Branchina:2024xzh, Branchina:2025hen}.)

At the heart of the Wilsonian EFT picture is the separation between ultraviolet (UV) and infrared (IR) physics~\cite{Appelquist:1974tg} (see also~\cite{Donoghue:2009mn}). 
This separation is enormously successful when the spacetime geometry can be treated as an external classical background. But gravity makes the situation conceptually different. The IR geometry is itself dynamical and reacts back to the energy carried by quantum fields. This raises the question whether the UV scale characterizing the quantum theory and the IR scale characterizing the geometry should be specified independently. The cosmological constant problem is the most direct manifestation of this tension: the vacuum contribution is computed as if the UV sector was independent of the size and geometry of spacetime, while the result is then used as a source for this same geometry. The hierarchy problem has a similar UV sensitivity, although it is usually discussed as a separate question. We explore the possibility that these two problems have the same structural origin, namely the assumption that the UV and IR sectors can be separated before gravity is taken into account~\cite{Berglund:2022qcc}.

  Another aspect of the problem relates to background independence.  A Wilsonian cutoff is usually introduced as a dimensionful scale which separates the modes that are kept from the ones that are integrated out~\cite{Wilson:1973jj, Wilson:1974mb, Wilson:1983xri, Weinberg:1978kz, Polchinski:1983gv}. A dimensionful quantity by itself is perfectly consistent, and a cutoff can be introduced in a diffeomorphism covariant way, for example through proper-time or heat-kernel methods~\cite{Vassilevich:2003xt}. The problem we want to stress here is rather that of background independence. A fixed dimensionful scale has to be measured with respect to some metric, while in gravity the metric is a dynamical field. If this scale is meant to have a physical meaning, it therefore represents an additional structure which is kept fixed while the geometry is varied. This is precisely the point at which the usual Wilsonian picture becomes less natural and where the introduction of what we call in this paper a Cardinal Renormalization parameter is more appropriate as it provides a relation between General Relativity and the Renormalization Group (in a different form compared to the relation between the boundary Renormalization Group and the bulk gravitational dynamics, found in the holographic context)~\cite{Maldacena:1997re, Gubser:1998bc, Witten:1998qj, Susskind:1998dq, deBoer:1999tgo}.
 It is also worthwhile to notice that  dimensional regularization respects covariance by avoiding introducing a physical cutoff scale. What it introduces instead is a reference scale in agreement with the Gell-Mann--Low--Stueckelberg--Petermann perspective on RG which differs from Wilson ~\cite{Gell-Mann:1954yli,Stueckelberg:1952hma,Bogolyubov:1959bfo,Callan:1970yg,Symanzik:1970rt}, and one studies how renormalized couplings and correlation functions depend on the normalization scale at which the renormalization
conditions are imposed. The mass scale is viewed as a reference, and not as a cut-off scale~\cite{Weinberg:1995mt,Weinberg:1996kr,Schwartz:2014sze}.




Our proposal starts from a different choice of what should be kept fixed when changing the scale.
Rather than separately
prescribing an infrared spacetime size and an ultraviolet momentum threshold, we restrict the quantum
theory to the \emph{accessible part of phase space} and characterize the size of this region by a
dimensionless number $\mathcal N$, which we call its \emph{cardinality}.
Our definition of $\mathcal N$ is semiclassical and follows directly from phase-space Weyl counting.
Let $\Omega_{\mathcal N}\subset T^*M$ denote an accessible region of off-shell
phase space. Its cardinality is the phase-space volume in units of elementary
cells. In the product approximation used below,
\begin{equation}
\mathcal N
=\frac{\operatorname{Vol}(\Omega_{\mathcal N})}{(2\pi\hbar)^d}
=\frac{\mathcal V\widetilde{\mathcal V}}{(2\pi\hbar)^d},
\label{intro:cardinality}
\end{equation}
where $\mathcal V$ and $\widetilde{\mathcal V}$ are the $d$-dimensional spacetime
and momentum-space volumes. This is semiclassical Weyl counting; $\mathcal N$
is not assumed to be an exact Hilbert-space dimension. Holding $\mathcal N$
fixed requires a change in one volume to be compensated by the other. We
evaluate the one-loop action of a scalar of mass $m$ through the corresponding
restricted trace at leading Weyl order. The elementary momentum integral is the same as in
cutoff field theory; the prescription changes what is held fixed when its upper
boundary varies.

One of the first remarkable effects of this assumption is that the classical gravitational action then participates in the RG equations. However,
the scale variation in what we call Cardinal RG=GR below has a different meaning compared to the usual RG equations. It
is a variation of a momentum scale while a dimensionless phase-space cardinality is held fixed, so
that the spacetime geometry varies with it.
We make this relation explicit on a round Euclidean de Sitter four-sphere of
radius $L$, with accessible momenta forming a four-ball of radius $\widetilde L$.
If $\Gamma$ denotes the dimensionless gravitational action plus the scalar
one-loop contribution, we have 
\begin{equation}
L^4\widetilde L^4=12\mathcal N\hbar^4,
\qquad
\left.L\frac{\mathrm d\Gamma}{\mathrm dL}\right|_{\mathcal N}
=-\left.\widetilde L\frac{\mathrm d\Gamma}{\mathrm d\widetilde L}
\right|_{\mathcal N}.
\label{intro:scale-variation}
\end{equation}
Thus geometric stationarity is equivalent to stationarity with respect to the
momentum scale. We call this relation Cardinal RG=GR. Throughout, Cardinal RG refers to this
constrained change of scale on the one-parameter de Sitter family, with the
geometry varying alongside the momentum domain. The central dynamical question
is whether the combined action has a stationary point and how the associated
curvature depends on the scalar mass.

The classical gravitational parameters define two useful quantities:
\begin{equation}
\mathcal N_G=S_{\rm GH}=\frac{3\pi}{\hbar G\Lambda},
\qquad
\varepsilon_G^4=\frac{4\pi\hbar^3\Lambda}{G}.
\label{intro:gravitational-scales}
\end{equation}
Here $G$ is Newton's constant, $\Lambda>0$ is the cosmological constant in the
classical action, and $S_{\rm GH}$ is the dimensionless Gibbons--Hawking entropy
of the classical de Sitter saddle~\cite{Gibbons:1976ue,Gibbons:1977mu}.
The scale $\varepsilon_G$ will govern the matter corrections, and with the gravitational
parameters used below, $\varepsilon_G\simeq9.4\times10^{-3}\,\mathrm{eV}$.
We distinguish the gravitational parameter $\mathcal N_G$ from the accessible
phase-space cardinality $\mathcal N$, leaving their relation to be determined
by additional physical input.
We assume a common cardinality for the gravitational and matter contributions and
we identify the entropy with minus the
on-shell action, $S=-\Gamma_\star$, and we utilize the bound
$0<\mathcal S\leq\mathcal S_{GH}$, where $S_{GH}$ is the corresponding Gibbons-Hawking entropy \cite{Gibbons:1976ue}, \cite{Gibbons:1977mu}.
Then we show that the cosmological constant, whose value is set by the above Gibbons-Hawking entropy of de Sitter space,
is radiatively stable.

We next couple a homogeneous scalar $\phi$ to a quantum scalar $\psi$ of mass
$m$ through the interaction $\lambda\phi^2\psi^2/2$, with $\lambda>0$. As in the case of the cosmological constant
we show that the scalar mass is radiatively stable.
In particular, the one-loop correction to the homogeneous scalar mass parameter is bounded by the gravitational scale $\varepsilon_G$, and decreases with the mass of a heavy field. (However, the extension to the physical Higgs pole mass requires a calculation of the momentum-dependent two-point function, including the full Standard Model particle content.)

Another important result is that conventional QFT emerges by considering a
correlated limit in which gravity decouples and the background becomes flat,
while the momentum scale governing the matter loop cutoff stays finite. This can be achieved by taking the double scaling  limit
$G \to 0,\Lambda\to0$ that keeps  ratios $\Lambda/G$ and $\mathcal{N}/\mathcal{N}_G$ fixed.
Note that in conventional
QFT, defined with a cutoff, the cutoff scale is  prescribed independently of any physical parameters and can be chosen to be arbitrarily large.  In our treatment this scale 
emerges from the expression for the vacuum energy in the flat space $\Lambda \to 0$
and non-gravitational limit $G \to 0$.
Another important point is that the ratio of the Wilsonian cut-off 
and the reference RG scale 
is given in terms of the ratio of the cardinal $\mathcal N$ and the gravitational entropy $\mathcal N_{G}$ associated with the cosmological horizon. This might indicate that quantum field theory (even in curved spacetime) should be defined, in terms of gravitational (and quantum gravitational) data, in a fully covariant and background independent manner.

The phase-space viewpoint is motivated by modular descriptions of quantum
spacetime~\cite{Freidel:2015pka,Freidel:2016pls} and by our earlier work linking
vacuum energy to gravitational entropy~\cite{Freidel:2022ryr,Freidel:2023ytq}.
Related constructions based on counting modes in gravity-coupled systems were
developed independently by Becker and
Reuter~\cite{Becker:2020mjl,Becker:2021pwo} and have since been investigated in
connection with vacuum energy and scalar effective
potentials~\cite{Ferrero:2024yvw,Branchina:2024xzh,Branchina:2025hen}.
The implementations differ, but they share the principle that geometry and the
counting entering the quantum theory are determined together.

{To summarize, the main points established in this work are: 1) the demonstration that the cosmological constant scales as $1/\sqrt{{\mathcal N}}$, where $\mathcal N$ is the cardinal number directly related to the volume of phase space, 2) the demonstration that the entropy scales as inverse of the cosmological constant and 3) the realization that the Wilsonian RG scale and the reference RG scale, as well as cosmological constant (or vacuum energy) are all functions of $\mathcal N$, providing a new view on Renormalization Group and the relation between Quantum Field Theory and gravity. Along the way we also establish radiative stability not only of the cosmological constant but of the mass of a scalar (``Higgs'') field as well.}

The paper is organized as follows: Section~II defines the cardinality and evaluates the scalar one-loop action.
Sections~III and~IV compare the scale variations and
introduce the gravitational dynamics.
Sections~V and~VI analyze the quantum corrected de Sitter 
solution and the entropic bound, 
while
Secs.~VII and~VIII discuss 
stability of the cosmological constant and the scalar (Higgs) field mass, respectively.\
In section~IX we demonstrate that curvature corrections do not change the central results of our work.
Appendices~A and~B give the spectral counting comparison and details
of the one-loop integration.

\section{Cardinality and the one-loop effective action}

 The discussion of the previous section suggests that the relevant restriction should be formulated not in terms of an externally prescribed momentum cutoff, but in terms of a finite accessible region of phase space. In this work we refer to the number of elementary phase-space cells contained in this region as the \emph{cardinality} and denote it by $\mathcal N$. It characterizes the combined size of the spacetime and momentum-space regions, without prescribing their scales independently. We study the consequence of finite cardinality on a matter system coupled to gravity, and focus on  what this implies for the cosmological constant and the stability of the  Higgs mass in a toy model. We now make the cardinality notion precise.

\subsection{Cardinality of the accessible phase-space region}

Let $M$ be a $d$-dimensional Euclidean spacetime and let $T^*M$ denote the corresponding phase space.
 This is the phase space attached to spacetime itself, and the natural space for the \emph{off-shell} variables entering the Euclidean functional trace. It should not be confused with the physical phase space of on-shell excitations associated with data on a codimension-one slice $\Sigma\subset M$.

The connection between the geometry of phase space and quantum theory is provided by Weyl quantization~\cite{Curtright:2011vw} and its generalization \cite{derezinski2020pseudodifferential}.
 The essential ingredient of this connection is the Weyl quantization map $Q_{\text{W}}:F\to\hat F$. For sufficiently regular symbols, it maps square-integrable phase-space functions to Hilbert--Schmidt operators, and real functions to symmetric operators. One of its key properties is that the trace pairing of two operators $(\hat F,\hat G)$ with Weyl symbols $F(q,p)$ and $G(q,p)$ is given by a phase-space integral
\begin{align}
 \Tr\left(\hat G^\dagger \hat F \right)
=
\int_{M\times \mathbb{R}^d}
\frac{ \rd \nu_q\, \rd^d p}{(2\pi\hbar)^d}\,
[\bar{G} {F} ](q,p),
\label{WeylTrace}
\end{align}
where $\rd\nu_q= \sqrt{g}(q) \rd ^d q$ is the Riemannian measure.
This construction is well known when $M=\mathbb{R}^d$ is flat. Quite remarkably, the trace formula can be extended to  curved Riemannian manifold without any correction \cite{derezinski2020pseudodifferential, andersson2025pseudodifferential}. The correspondence is  referred to in this case as  balanced geodesic Weyl quantization.
 Given an operator $\hat F$, we denote its integral kernel by $\hat F(q,q')=\langle q|\hat F|q'\rangle$. Its balanced Weyl symbol is written as
\be \label{Weyl}
F(q,p) = \int_{\mathbb{R}^d} \frac{\hat{F}(q[-u] ,q[+u])}{\Upsilon(q[-u] ,q[+u])}  e^{i u\cdot p/\hbar}   \rd^d u.
\ee
Here $q[\tau u]=\exp_q(\tau u/2)$ denotes the geodesic starting at $q$ with tangent vector $\tau u/2\in T_qM$.
The factor $\Upsilon(x,y)$ accounts for the Jacobian
of the transformation from the endpoints $(x,y)$
to the geodesic midpoint $q$ and separation
$u\in T_qM$. With metric-induced measures and
the corresponding scalar-kernel convention,
it satisfies
\begin{equation}
\Upsilon^2(x,y)\,d\nu_x\,d\nu_y
= d\nu_q\,d\mu_q(u),
\qquad
\Upsilon(x,y)=\Delta^{1/2}(x,y),
\end{equation}
where $\Delta(x,y)$ is the Van Vleck–Morette determinant.

 The main idea developed in \cite{Freidel:2022ryr, Freidel:2023ytq} is that quantum gravity should lead to a physical restriction of the phase space accessible to quantum excitations. We implement this prescription by postulating that only a finite region $\Omega_{\mathcal N}\subset T^*M$ is accessible. The restriction is described by its characteristic function $\chi_{\mathcal N}(q,p)$, which equals one inside $\Omega_{\mathcal N}$ and vanishes outside. It satisfies $\chi_{\mathcal N}^2=\chi_{\mathcal N}$. Multiplication by $\chi_{\mathcal N}$ thus defines a projection on phase-space functions.

The Weyl formula \eqref{WeylTrace} shows that the natural elementary volume in a $2d$-dimensional phase space is $(2\pi\hbar)^d$. We therefore define the \emph{cardinality} of the accessible region to be the number of such elementary cells contained in $\Omega_{\mathcal N}$,
\begin{align}
\mathcal N
:=
\int_{T^*M}
\frac{\rd\nu_q\,\rd^d p }{(2\pi\hbar)^d}\,
\chi_{\mathcal N}(q,p)
=
\frac{\operatorname{Vol}(\Omega_{\mathcal N})}{(2\pi\hbar)^d}.
\end{align}
 We  treat $\mathcal N$ as a semiclassical phase-space counting quantity, and for large $\mathcal N$ also as a continuous variable.  The term \textit{cardinality} is meant to emphasize that the basic datum is this dimensionless number, rather than a dimensionful scale specifying separately the extent of the accessible region in position or momentum space.

 The phase-space volume in unit of $h$ is the definition of $\mathcal N$ used here. The balanced Weyl trace pairing may be defined on curved spaces, but its identification with a spectral count is in general valid only at leading semiclassical order. We will return to the subleading geometric corrections below. In this section we use this leading approximation.
 In local canonical coordinates the integral may be decomposed into elementary cells $C(k,\tilde k)$ of symplectic volume $(2\pi\hbar)^d$,
\begin{align}
\int_{T^*M}
\frac{\rd^d p\,\rd^d q}{(2\pi\hbar)^d}\,
F(q,p)
=
\sum_{(k,\tilde k)\in\mathbb Z^{2d}}
\int_{C(k,\tilde k)}
\frac{\rd^d p\,\rd^d q}{(2\pi\hbar)^d}\,
F(q,p).
\end{align}
This phase-space description is closely related to the Zak-transform and modular-variable formulation~\cite{aharonov2008quantum, Freidel:2016pls}.

Correspondingly, we define the trace associated with the accessible region by
\begin{align}\label{traceid}
\Tr_{\mathcal N}\left(\hat F\right)
:= \Tr\left(\hat{\chi}_N \hat F\right)=
\int_{T^*M}
\frac{\rd^d p\,\rd \nu_q}{(2\pi\hbar)^d}\,
\chi_{\mathcal N}(q,p)F(q,p)
=
\int_{\Omega_{\mathcal N}}
\frac{\rd^d p\,\rd \nu_q}{(2\pi\hbar)^d}\,
F(q,p),
\end{align}
where the equality involve the Weyl operator $\hat{\chi}_N$ associated with the classical projector.\footnote{ While $\chi_{\mathcal N}^2=\chi_{\mathcal N}$ at the classical level,  this is no longer true at the quantum level: $\hat{\chi}_N^2\neq \hat{\chi}_N $ due to the presence of non-commutative correction in the star product. } For $\hat F=\mathbb I$ one recovers the cardinality $\mathcal N$\footnote{ If the manifold $M$ is not
parallelizable, as is the case for $S^4_L$, the accessible
phase-space region should be defined as a cotangent ball
bundle over $M$, rather than as a Cartesian product.
This distinction does not affect the calculations below
at leading Weyl order.}.
For the applications below we take the accessible region to have the product form $\Omega_{\mathcal N}=\tilde M\times M$, where $M$ is assumed for simplicity to be a compact spacetime of finite volume $\mathcal V$ and $\tilde M$ is a region in momentum space of volume $\tilde{\mathcal V}$. The definition of cardinality then becomes
\be \label{CardR}
\frac{\mathcal{V}\tilde{\mathcal{V}}}{(2\pi\hbar)^d}=\mathcal N =\Tr(\hat{\chi}_N).
\ee
The relation \eqref{CardR} fixes the product of the spacetime and momentum-space volumes. The connection between $\mathcal N$ and  quantum state count is what we now investigate.
\subsection{Cardinality and spectral counting}

 The Cardinality $\mathcal{N}$ can be related to the counting of eigenvalues of a Laplace-type operator in the limit of large cardinality.
  To see this let us consider a  diffeomorphism- and gauge-covariant Laplace-type operator acting on a scalar multiplet,
\be
 P= -g^{ab} D_a D_b + U, \qquad D_a:=(\nabla_a + A_a ),
\ee
where $\nabla_a$ is the gravitational covariant derivative compatible with the metric $g$, $U$ is a potential term and $D_a$ is the covariant derivative of a Hermitian bundle of rank $N_E$.
 The operator $P$ has dimensions of inverse length squared. We assume it to be non-negative for the spectral counting discussed here. In agreement with Appendix~\ref{app:A}, we introduce a momentum scale\footnote{We use the notation  $\tL$ for the momentum scale instead of the traditional $\Lambda$ because $\Lambda$ denotes the cosmological constant and $\tL$ refers to the dual length scale on par with the cosmological scale $L$.} $\tL$ and consider the spectral projector $\Pi_P({\tL}):= \Theta(\tL^2 - \hbar^2 P) $ which projects out all the eigenvalues $\lambda_i$ of $\hbar^2 P$ that are above the scale $\tL^2$ and define the sharp spectral count
\be
N_{\rm sharp}(\tL) := \Tr \Pi_P(\tL).
\ee
The Weyl formula \cite{Ivrii:2016} implies that in the limit of large cutoff $\tL$ the spectral number asymptotes the cardinality times the number of fields
\be
N_{\rm sharp}(\tL) \sim N_E \frac{\mathcal{V}_M\tilde{\mathcal V}_{\tL}}{(2\pi \hbar)^d} = N_E \mathcal{N},
\ee
where $\mathcal{V}_M=\int_M \rd\nu_x$ denotes the spacetime volume and  $\tilde{\mathcal V}_{\tL}$ is the  volume of a $d$-dimensional momentum ball of radius $\tL$
\be
\tilde{\mathcal V}_{\tL}=
\frac{(2\pi\tL)^d}{(4\pi)^{d/2}\Gamma(d/2+1)}.
\ee
In order to go beyond this asymptotic evaluation and evaluate the subleading local terms we need to replace the sharp cut-off by a smooth cutoff function $F$ that is in the image of the Laplace transform\footnote{The Paley-Wiener theorem describing the Hardy space.} and which is such that $F(0)=1$ while it exponentially decays near $x\to \infty$ (see Appendix \ref{app:A}).
In order for this smooth approximation to reproduce the local dimension we impose that the first few moments of $F$ are identical to the moments of the hard cut-off, which means that
\be
F_0=F_1=1, \quad F_2=\frac12
\ee
where the moments are
$
F_n := -  \int_0^\infty \frac{x^n}{n!}  F'(x) \rd x.$
In particular we have that $F_0=F(0)=1$. These moments conditions provide the local approximation to the sharp spectral count.

Under these conditions we find in four dimensions, as explained in Appendix~\ref{app:A}, a universal behavior of the first three dominant  terms~\cite{Chamseddine:2009drp,Chamseddine:2010ud, Chamseddine:2012sw}
$N_F(\tL)= N_P^{\rm loc}(\tL) + O(1/\tL^2)$ where
\be
N_P^{\rm loc}(\tL)
=\frac{N_E}{(4\pi)^2}\left[
\frac{\tL^4}{2\hbar^4}\,\mathcal V_M
-\frac{\tL^2}{\hbar^2}\int_M W\,\rd\nu_x
+\int_M C\,\rd\nu_x
\right].
\ee
The terms $W$ and $C$ are local functionals that are determoned by the heat kernel expansion and given by
\be
 W:=\tau_E(W_E),\qquad  W_E:=U-\frac16R\,\mathbb I_E,
\ee
where $\tau_E(\cdot):=\tfrac1{N_E}\tr_E(\cdot)$ is the normalised trace,
and
\be C:=\frac12\tau_E(W_E^2)+\frac1{12}\tau_E(F_{\mu\nu}F^{\mu\nu})
+\frac1{180}\left(R_{\mu\nu\alpha\beta}R^{\mu\nu\alpha\beta}-R_{\mu\nu}R^{\mu\nu}\right).
\ee
 Using the relation
$
\mathcal N=\frac{\mathcal V\tL^4}{32\pi^2\hbar^4}
$,
 for a four-dimensional momentum ball, we see that the first term is the Weyl term $N_E\mathcal N$. The local curvature contributions give the subleading terms of the smoothed count. In dimension four we can write this expansion in terms of $\mathcal N$ as follows:
\begin{equation}
\frac{N_P^{\rm loc}(\mathcal N)}{N_E}
\sim \mathcal N-A_F(g,A)\sqrt{\mathcal N}+B_F(g,A).
\label{eq:N-d4}
\end{equation}
 The coefficients are given by
\be
A_F(g,A)= \frac{1}{\sqrt{8\pi^2\mathcal V}}\int_M W\,\rd\nu_x,
\qquad
B_F(g,A)=\frac{1}{16\pi^2}\int_M C\,\rd\nu_x.
\ee
 They are functionals invariant under global rescaling $g\to L^2 g$ which measures the rescaling invariant geometry encoded in $P(g,A)$.  This shows that the local spectral counting formula is directly connected to the cardinality $\mathcal{N}$ with subleading corrections due to the curvature and potential terms.

Equation~\eqref{CardR} is the relation that will be used throughout the paper. It expresses the size of the accessible phase-space region in units of the elementary cell $(2\pi\hbar)^d$. In this form it also reproduces the leading Weyl counting that appears in spectral approaches to quantum geometry and in non-commutative geometry~\cite{Chamseddine:2009drp,Chamseddine:2010ud, Chamseddine:2012sw}.

\subsection{The one-loop effective action}

We now turn to the one-loop effective action of a free massive scalar field. The trace entering the effective action is understood from the outset as the  geometrically restricted trace associated with the accessible region $\Omega_{\mathcal N}$ defined above. As detailed in the previous section, it corresponds to the leading contribution to the spectral dimension.  After Wick rotation we write
\be \label{Gamma1}
\Gamma_1
:=
\frac12\Tr_{\mathcal N}\ln\!\left[\mu^{-2}(\Box_g+m^2)\right]
=
\frac12\int_0^{\infty}\frac{\rd\tau}{\tau}\,Z(m;\tau),
\ee
where  $\Box_g :=-\hbar^2\nabla^a\nabla_a$ is the positive covariant Laplace--Beltrami operator and indices are contracted with the Euclidean metric\footnote{We will denote the flat-space kinetic operator by $\Box=-\hbar^2\delta^{ab}\pa_a\pa_b$.}. The second equality is the Schwinger representation, with
\be \label{partf}
Z(m;\tau)
:=
\Tr_{\mathcal N}\left(
e^{-\mu^2\tau}
-e^{-m^2\tau}e^{-\tau\Box_g}
\right).
\ee
The scale $\mu$ is required because the argument of the logarithm must be dimensionless. It is a normalization scale of the determinant and should not be confused with the scales characterizing the shape of the accessible phase-space region $\Omega_{\mathcal N}$.

Let us recall a small but important point concerning the Schwinger representation. One often writes formally
\begin{align}\label{lnAintegral}
\ln A=-\int_0^\infty\frac{\rd\tau}{\tau}\,e^{-\tau A}.
\end{align}
As it stands, this expression contains a divergence at $\tau=0$. The appropriate identity for the dimensionless logarithm is instead
\begin{align}\label{reglnA}
\ln\left(\frac{A}{\mu^2}\right)
=
\int_0^\infty\frac{\rd\tau}{\tau}
\left(
e^{-\mu^2\tau}\mathbb I
-e^{-\tau A}
\right).
\end{align}
The subtraction proportional to the identity is not an ultraviolet regulator. It is a reference scale  required to define the logarithm of a dimensionful positive operator.

The choice of normalization becomes relevant when the absolute value of the effective action is used, as will be the case later in the entropy discussion. The dependence of the functional measure on such a normalization has been discussed, in particular, in connection with the Fradkin--Vilkovisky prescription~\cite{fradkin1973s, Branchina:2025lqw}. In the present paper we adopt the prescription
\be \label{epsilon=mu} \mu^2=m^2+\varepsilon^2,
\ee
 Here $\varepsilon$ is a reference scale whose role will be discussed in Sect. \ref{Sect:EntropyBound}.
 We regard \eqref{epsilon=mu} as a choice of normalization, separate from the definition of the cardinality $\mathcal N$. It has the useful property that for $m\gg\varepsilon$ one has $\mu\simeq m$, whereas for a massless field $\mu=\varepsilon$. The form of \eqref{epsilon=mu} is also reminiscent of the dispersion relation encountered in metaparticle dynamics based on Born geometry~\cite{Freidel:2018apz}.


\subsection{Effective action evaluation in four dimensions}

To evaluate the one-loop effective action we take the Euclidean spacetime to be a compact Riemannian manifold $M$. In the gravitational application below it will be a four-sphere $S^4_L$ of radius $L$, obtained by Euclidean continuation of de Sitter space. We choose the momentum-space part of the accessible region, $\tilde M$, to be a four-dimensional ball $B_4$ of radius $\tL$. From \eqref{CardR} one then has
\be \label{volumens}
\mathcal{V}_{S_4}=\frac{8\pi^2L^4}{3},\qquad
\tilde{\mathcal{V}}_{B_4}=\frac{\pi^2\tL^4}{2},\qquad
L^4\tL^4=12\mathcal N\hbar^4.
\ee
The leading Weyl description is sufficient for the large-$\mathcal N$, weak-curvature regime considered here. Curvature-dependent corrections are subleading in this regime and will be briefly discussed in the conclusion.  The product form $M\times\tilde M$ used here is to be understood in this leading phase-space approximation.

Writing the trace in \eqref{Gamma1} explicitly and using the normalization prescription \eqref{epsilon=mu}, we obtain
\begin{align}\label{Gamma2}
\Gamma_1
&=\frac12\int_0^{\infty}\frac{\rd\tau}{\tau}
\int_{M\times\tilde M}\frac{ \rd\nu_q\,\rd^4p }{(2\pi\hbar)^4}\,
 e^{-m^2\tau}
\left[e^{-\varepsilon^2\tau}-e^{-p^2\tau}\right]
\nonumber\\
&=\frac{\mathcal N}{2}\int_0^{\infty}\frac{\rd\tau}{\tau}
\int_{\tilde M}\frac{\rd\nu_p}{\tilde{\mathcal V}}\,
 e^{-m^2\tau}
\left[e^{-\varepsilon^2\tau}-e^{-p^2\tau}\right],
\end{align}
where $p^2=\delta^{ab}p_ap_b$. This is the leading Weyl representation of the trace over the accessible phase-space region.

For a four-dimensional momentum-space ball of radius $\tL$ one finds
\begin{align}
Z_{\mathcal N}(\tau)
&=\frac{4\mathcal N}{\tL^4}
\int_0^{\tL}p^3\rd p\,e^{-m^2\tau}
\left[e^{-\varepsilon^2\tau}-e^{-p^2\tau}\right]
\nonumber\\
&=\mathcal N e^{-m^2\tau}
\left[
 e^{-\varepsilon^2\tau}
 -\frac{2}{\tL^4\tau^2}
 +2\frac{e^{-\tL^2\tau}}{\tL^4\tau^2}
 \left(1+\tL^2\tau\right)
\right].
\end{align}
The individual terms in the proper-time integral are singular at $\tau=0$, but their sum is finite as a consequence of the subtraction in \eqref{reglnA}. The remaining integral is a generalized Frullani integral and is evaluated exactly  in Appendix \ref{AppA}. The result is that
\be \label{1loopaction}
\Gamma_1= \mathcal N \gamma_1,
\ee
 so that the effective action is \emph{extensive} in the cardinality $\mathcal N$. The dependence on the momentum-space radius only enters the reduced one-loop effective action $\gamma_1$, which is given by
\begin{align}
\label{Gammaex}
\boxed{
\gamma_1=
\frac{ 1}{2}
\left[
\ln\left(\frac{\tL^2+m^2}{\varepsilon^2+m^2}\right)
-\frac12
+\frac{m^2}{\tL^2}
-\frac{m^4}{\tL^4}
\ln\left(\frac{\tL^2+m^2}{m^2}\right)
\right]
}.
\end{align}
This is the one-loop expression that will be used in the following sections.
 Its dependence on the momentum-space radius is only logarithmic for $\tL\gg m$,
\begin{align}
\gamma_1
=\frac{1}{2}
\left[
\ln\left(\frac{\tL^2}{\varepsilon^2+m^2}\right)
-\frac12
+\mathcal O\!\left(\frac{m^2}{\tL^2}\right)
\right]\,,\quad \tL \gg m
\end{align}
where the omitted terms include contributions of order
$\mathcal O\!\left(m^4\tL^{-4}\ln(\tL^2/m^2)\right)$.
 The striking property of the effective action
 $\Gamma_1= N \gamma_1$ is that at fixed $\mathcal N$ it contains no quartic or quadratic growth when $\tL\to\infty$. In this limit the spacetime volume decreases according to \eqref{CardR}. For fixed cardinality the dependence is purely logarithmic. This is the key property underlying the mechanism discussed below for the cosmological constant and hierarchy problems. This behavior should not be described as the regularization of a quartically divergent expression. Rather, it is a property of the effective action evaluated over an accessible phase-space region of fixed cardinality $\mathcal N$.
In the opposite limit, for $m\neq0$,
\begin{align}
\lim_{\tL\rightarrow0}\Gamma_1
=
\frac{\mathcal N}{2}
\ln\left(\frac{m^2}{m^2+\varepsilon^2}\right).
\end{align}
The non-zero limiting value reflects the normalization prescription \eqref{epsilon=mu}.

The structural point that will be important below is the relation \eqref{CardR}. At fixed cardinality, the spacetime and momentum-space volumes entering the accessible region cannot be varied independently. Thus varying $\tL$ at fixed $\mathcal N$ is different from varying a momentum cutoff at fixed spacetime volume. This distinction will be made explicit in the comparison with the standard EFT expression.

 Let us note that closely related finite phase-space counting constructions arise in modular quantization \cite{Freidel:2016pls,Freidel:2022ryr, hubsch2024quantum}. Related ideas have been developed by Becker and Reuter in \cite{Becker:2020mjl,Becker:2021pwo} and further developed by Ferrero and Percacci in \cite{Ferrero:2025ugd,Ferrero:2024yvw}.  A similar idea has been utilized in \cite{Cadoni:2026qss}. The works of Branchina et al.\ \cite{Branchina:2024xzh,Branchina:2025hen,Branchina:2026vvk,Branchina:2024lai} are also relevant for the new concept of cardinal regularization.

\section{Fixed volume versus fixed cardinality}\label{Sect:FixedN}

The result obtained in the previous section can be directly compared with the standard effective field theory calculation. This comparison is useful because it makes clear that the difference between the two descriptions does not originate in a different evaluation of the momentum integral. Rather, it comes from what is kept fixed when the momentum-space scale $\tL$ is varied. In the usual EFT treatment the spacetime volume $\mathcal V$ is fixed, whereas in the construction considered here the cardinality $\mathcal N$ of the accessible phase-space region is fixed.

\subsection{Comparison with EFT}

Let us first recall the standard cutoff calculation. The one-loop vacuum energy density of a free massive scalar field, evaluated with an Euclidean momentum cutoff $|k_E|<\tL$, is
\begin{equation}
\rho_{\tilde L}^{(1)}(m;\mu)
:=
\frac{1}{2}
\int_{\lvert k_E\rvert<\tilde L}
\frac{d^4k_E}{(2\pi)^4}
\ln\!\left(
\frac{k_E^2+m^2}{\mu^2}
\right),
\label{eq:V1_cutoff}
\end{equation}
where, as in the previous section, the normalization scale $\varepsilon$ is introduced in order to make the argument of the logarithm dimensionless. Using $d^4k=2\pi^2k^3\,dk$, we obtain
\begin{align}
\rho_{\tilde L}^{(1)}(m;\mu)
&=
\frac{1}{16\pi^2}
\int_0^{\tilde L}
k^3
\ln\!\left(
\frac{k^2+m^2}{\mu^2}
\right)dk
\nonumber\\
&=
\frac{1}{32\pi^2}
\int_0^{\tilde L^2}
z
\ln\!\left(
\frac{z+m^2}{\mu^2}
\right)dz,
\label{eq:radial1}
\end{align}
where $z=k^2$, so that $k^3dk=\tfrac12 z\,dz$. The required integral is
\begin{equation}
\int_0^x
z\ln\!\left(
\frac{z+m^2}{\mu^2}
\right)dz
=
\frac12
\left[
x^2
\ln\!\left(
\frac{x+m^2}{\mu^2}
\right)
-
m^4
\ln\!\left(
\frac{x+m^2}{m^2}
\right)
+
m^2x
-
\frac{x^2}{2}
\right].
\label{eq:antideriv}
\end{equation}
Applying this to Eq.~\eqref{eq:radial1} gives
\begin{equation}\label{vace}
\rho_{\tilde L}^{(1)}
=
\frac{1}{64\pi^2}
\left[
\tilde L^4
\ln\!\left(
\frac{\tilde L^2+m^2}{\mu^2}
\right)
-
m^4
\ln\!\left(
\frac{\tilde L^2+m^2}{m^2}
\right)
+
m^2\tilde L^2
-
\frac{\tilde L^4}{2}
\right].
\end{equation}
For large $\tL$ this becomes
\begin{equation}
\begin{split}
\rho_{\tilde L}^{(1)}
=
\frac{1}{64\pi^2}
\Bigg[
& -
\frac{\tilde L^4}{2} +
2m^2\tilde L^2 +
\tilde L^4
\ln\!\left(
\frac{\tilde L^2}{\varepsilon^2}
\right)-
m^4
\ln\!\left(
\frac{\tilde L^2}{m^2}
\right)\Bigg] + O(1).
\end{split}
\label{eq:V1_cutoff_expand}
\end{equation}
The quartic and quadratic terms are the familiar power sensitivities of the vacuum energy, and of scalar masses, to the ultraviolet scale in a cutoff/Wilsonian description.

Using the expression for the vacuum energy  together with Eq.~\eqref{Gammaex}, we find
\be
\rho_{\tilde L}^{(1)}(m;\mu)
=
\frac{\tilde L^4}{32\pi^2}
\left[
\gamma_1(\tilde L)
+
\frac{ 1}{2}
\ln\!\left(
\frac{\varepsilon^2+m^2}{\mu^2}
\right)
\right].
\label{eq:rhoGammaRelation}
\ee
If the normalization is chosen as in Eq.~\eqref{epsilon=mu}, $\mu^2=m^2+\varepsilon^2$, the last term vanishes.

Thus the fixed-cardinality expression and the standard cutoff expression are two ways of organizing the same one-loop integral. The essential difference is not the momentum integration itself, but the constraint imposed when the scale $\tL$ is varied.
To understand this important point, it is essential to multiply the energy density by the spacetime volume and using Eq.~\eqref{CardR}, specialized to the four-sphere and the four-dimensional momentum ball, we have
\be
\frac{\mathcal{V}_{S^4}}{\hbar^4}
\frac{\tL^4}{64\pi^2}
=
\frac{L^4\tL^4}{24\hbar^4}
=
\frac{\mathcal N}{2}.
\label{eq:prefactorreplacement}
\ee
This means that eq.~\eqref{eq:rhoGammaRelation} reduces to
\be
\left.
\rho_{\tilde L}^{(1)}\right|_{\mu^2=m^2+\varepsilon^2}
=
\frac{\tilde L^4}{32\pi^2}
\gamma_1(\tilde L) = \frac{\mathcal{N}\hbar^4}{\mathcal{V}_{S^4}}
\gamma_1(\tilde L).
\label{eq:rhoGammaDirect}
\ee

 Therefore, at fixed spacetime volume $\mathcal V$, Eq.~\eqref{CardR} implies $\mathcal N\propto \tL^4$, meaning the usual $\tL^4$ behavior of the vacuum energy is due to the quartic growth of accessible phase-space cells. At fixed cardinality, on the other hand, because $\Gamma_1$ is a dimensionless quantity that is extensive in the cardinality $\mathcal N$, an increase of the momentum-space volume must be accompanied by a compensating decrease of the spacetime volume. Dividing by this volume to obtain an energy density reintroduces an explicit volume dependence, leading to the logarithmic large-$\tL$ behavior found in Eq.~\eqref{Gammaex}. In this sense, the two calculations differ by the condition under which the scale variation is performed.

\subsection{Scale variation at fixed cardinality}

This distinction can be stated directly in differential form. Let the effective action be regarded as a function $\Gamma(\mathcal N,\tL)$. At fixed spacetime volume the Weyl relation implies, in four dimensions,
\be
\left.\tL\frac{\rd\mathcal N}{\rd\tL}\right|_{\mathcal V}
=
4\mathcal N.
\label{eq:Nscaling}
\ee
Therefore the scale derivative at fixed spacetime volume is
\be
\left.\tL\frac{\rd\Gamma}{\rd\tL}\right|_{\mathcal V}
=
\left.\tL\pa_{\tL}\Gamma\right|_{\mathcal N}
+
4\mathcal N\,\pa_{\mathcal N}\Gamma.
\label{eq:fixedVvariation}
\ee
By contrast, the scale variation at fixed cardinality is simply
\be
\left.\tL\frac{\rd\Gamma}{\rd\tL}\right|_{\mathcal N}
=
\left.\tL\pa_{\tL}\Gamma\right|_{\mathcal N}.
\label{eq:fixedNvariation}
\ee
Equations~\eqref{eq:fixedVvariation} and \eqref{eq:fixedNvariation} make precise the distinction that will be important below. The same parameter $\tL$ can be varied in two inequivalent ways: at fixed background volume, in which case the cardinality changes, or at fixed cardinality, in which case the spacetime and momentum-space parts of the accessible region change together. A related distinction between a fixed number of modes and a fixed background cutoff has also been emphasized in Ref.~\cite{Branchina:2025hen}.


\section{Cardinal RG=GR}\label{Sect:GR=RG}

We now bring gravity into the discussion. The main point is that when we fix  $\mathcal N$ then the classical gravitational action depends on the momentum scale $\tL$, since a change of this scale must be accompanied by a change of the geometry size.
This means that the classical action  contributes to the renormalization group while it usually enters one through the one-loop matter corrections.

In the minisuperspace setting considered here the Euclidean spacetime is a four-sphere $S^4_L$ of radius $L$, while the momentum-space part of the accessible region is a four-dimensional ball $B^4_{\tilde L}$ of radius $\tilde L$. Their volumes are
$\mathcal{V}_{S^4_L}=8\pi^2L^4/3$ and $\widetilde{\mathcal V}_{B^4_{\tilde L}}=\pi^2\tilde L^4/2$, respectively, and the scalar curvature of $S^4_L$ is ${R}(g)=12/L^2$.

In Euclidean quantum gravity the path-integral weight is $e^{-\Gamma_0}$, where $\Gamma_0=S_E/\hbar$ is the dimensionless Euclidean Einstein action. In $d=4$ it is given by~\cite{Gibbons:1976ue,Gibbons:1977mu,Gibbons:1994cg}
\begin{align}\label{classactionx}
\Gamma_{0}(L)
&= \frac1{16\pi \hbar G}\int\rd^4x\sqrt{g}\qty(2\Lambda-{R}(g))
=\frac{\pi}{\hbar G}\left(\frac{\Lambda}{3}L^4-2L^2\right).
\end{align}

 The leading Weyl relation between the size of the accessible phase-space region and its cardinality is
\be \label{RtR}
L^4\tL^4=12\mathcal N\hbar^4.
\ee
In our framework $\mathcal N$ is fixed. Therefore the spacetime radius $L$ and the momentum-space radius $\tilde L$ are not independent: along the surface of fixed cardinality they are related one-to-one by Eq.~\eqref{RtR}. Consequently, within the present one-parameter minisuperspace, the stationarity condition obtained by varying the geometry with respect to $L$ can equivalently be written as a stationarity condition with respect to $\tilde L$.  In this sense the cardinal renormalization group flow  leads to the same equation as the Einstein equation for global changes of geometry and  we have
\begin{align}
\mathrm{Cardinal\,RG}=\mathrm{GR}.
\end{align}
The Cardinal Renormalization group which demands the invariance of the action under variation of the momentum-space scale at fixed cardinality, rather than to the conventional Wilsonian flow at fixed spacetime volume.

Using Eq.~\eqref{RtR}, the gravitational action can be written as a function of $\tilde L$,
\be \label{classaction}
\Gamma_0(\tL)
= \frac{\pi\hbar}{G}\left(
\frac{\Lambda}{3}\frac{12\mathcal N}{\tL^4}
-2\frac{\sqrt{12\mathcal N}}{\tL^2}
\right)
=\frac{\pi m_{\mathrm P}^2}{m_{\mathrm c}^2}\left(
\frac{12\mathcal N m_{\mathrm c}^4}{\tL^4}
-2\frac{\sqrt{12\mathcal N}m_{\mathrm c}^2}{\tL^2}
\right),
\ee
where we introduced the Planck mass $m_{\mathrm P}^2=\hbar/G$ and the cosmological mass $m_{\mathrm c}$, dual to the cosmological length $\ell_{\mathrm c}$,\footnote{Numerically we have ~\cite{Planck:2018vyg},
$
\ell_{\mathrm c}\simeq1.65\times10^{26}\,\mathrm m$
and
$m_{\mathrm c}\simeq1.19\times10^{-33}\,\mathrm{eV}.
$
}
\begin{align}
\ell_{\mathrm c}^2=\frac{3}{\Lambda},
\qquad
m_{\mathrm c}^2=\frac{\hbar^2\Lambda}{3}.
\end{align}

To analyze this action it is convenient to introduce a dimensionless gravitational parameter $\mathcal N_G$ and a vacuum energy scale $\varepsilon_G$, which are defined by
\be \label{UVIR}
\mathcal N_G:=\pi\frac{m_{\mathrm P}^2}{m_{\mathrm c}^2}\simeq 3.3 \times 10^{122}, \qquad
\varepsilon_G^2
:=\sqrt{12\pi}\,m_{\mathrm P}m_{\mathrm c} \simeq(9.44\times10^{-3}\,\mathrm{eV})^2.
\ee
The scale $\varepsilon_G$ is, quite remarkably, of the same general order as the fourth root of the observed dark-energy density, $\rho_\Lambda^{1/4}\simeq 2\,\mathrm{meV}$. We will refer to it as the vacuum energy scale. It is such that  $\mathcal N_G^{1/4}\varepsilon_G$ is of order the Planck mass.
 The action then reads
\be \label{classaction2}
\Gamma_0(\tL)
= \mathcal N \left(
\frac{\varepsilon_G^4}{\tL^4}
-2\sqrt{\frac{\mathcal N_G}{\mathcal N}}\frac{\varepsilon_G^2}{\tL^2} \right).
\ee
The cardinal RG equation imposes the stationarity of $\Gamma_0$ under variation of $\tL$ at fixed $\mathcal N$.
This  stationarity condition possesses two solutions. One is the singular solution $\tL=\infty$ which corresponds to a universe of zero size. The other solution has a finite value
\be \label{classEEsolusion}
\tL_*^2= \sqrt{\frac{\mathcal N}{\mathcal N_G}}\varepsilon_G^2
.
\ee
As is well established, the regular Euclidean de Sitter saddle has an entropy given by the Gibbons--Hawking entropy~\cite{Gibbons:1976ue}, which is minus the Euclidean action evaluated on shell,
\begin{align}\label{entm}
S_{\mathrm{GH}}=-\Gamma_0\big|_{\mathrm{on\ shell}} = \mathcal{N}_G.
\end{align}
The value of the gravitational entropy for the regular solution is  $S_{\mathrm{GH}}=\mathcal{N}_G$ while its value for the singular solution is $S_{\mathrm{GH}}=0$.

In this work we use the principle of maximum entropy  as a selection argument on which solution contributes. Since the regular solution has the larger entropy this means that universe with finite size is selected and we recover the  usual Gibbons--Hawking expression
\begin{align}
S_{\mathrm{GH}}=\frac{A_{\mathrm c}}{4\hbar G} =\mathcal{N}_G,
\end{align}
where $A_{\mathrm c}=4\pi\ell_{\mathrm c}^2$ is the area of the cosmological horizon.

The classical stationary solution determines the
de Sitter radius independently of the accessible
phase-space cardinality $\mathcal N$. The latter
determines the corresponding momentum-space scale
through Eq.~(51). The relation between $\mathcal N$
and the gravitational entropy $\mathcal N_G$ will
be investigated in the following sections, after
including the one-loop matter contribution.

Note finally that the scale $\varepsilon_G$  \eqref{UVIR} is given by  the geometric mean of the gravitational UV and IR scales.
This is the UV/IR relation associated with the fixed-cardinality RG=GR correspondence.
 This scale is comparable to the neutrino oscillation  scales  and below the upper bound on neutrino masses
\be \nonumber
\text{solar}:\,\,\frac{\sqrt{\Delta m_{21}^2}}{\varepsilon_G}  \approx 0.83, \quad \text{atmospheric}:\,\frac{\sqrt{|\Delta m_{32}^2|}}{\varepsilon_G}  \approx  4.76,\quad\text{upper\, bound}:
\frac{\sum m_\nu}{\varepsilon_G}  \approx 11.6.
\ee
It is therefore tiny compared with
all charged-lepton, baryon, quark, and Higgs masses.  A few striking ratio for the electron, muon, proton, Higgs and top quark are
\be\nonumber
\frac{m_e}{\varepsilon_G}\approx 4.92\times10^7,
\,\,
\frac{m_\mu}{\varepsilon_G}\approx 1.02\times10^{10},
\,\,
\frac{m_p}{\varepsilon_G}\approx 9.04\times10^{10},
\,\,
\frac{m_H}{\varepsilon_G}\approx 1.21\times10^{13},
\,\,
\frac{m_t}{\varepsilon_G}\approx 1.66\times10^{13}.
\ee

\section{Quantum-corrected de Sitter solution} 
\label{SectVSemiClass}

We now include the one-loop matter contribution and look for the stationary solution of the gravity--matter system.
The classical gravitational action is kept in the form obtained in Sec.~\ref{Sect:GR=RG},
and the matter one-loop contribution is given in eq \ref{1loopaction}.

In forming the total action we use that the cardinality $\mathcal N$ entering
the gravitational action is also the cardinality used to restrict the matter
one-loop trace. This follows simply by assuming that the classical action and one loop fluctuations relates to the same universal spacetime phase space.
 Notice that this assumption is distinct from identifying
$\mathcal N$ with the gravitational parameter $\mathcal N_G$ defined in
Eq.~\eqref{UVIR}. We keep these two quantities independent when looking for
the quantum-corrected stationary solutions, and return to their relation below.

Thus the total action $\Gamma=\Gamma_0 +\Gamma_1$ is
\begin{equation}
\begin{aligned}
\Gamma(\tL,\mathcal N;m)
={}&
\mathcal N \left(
\frac{\varepsilon_G^4}{\tL^4}
-2\sqrt{\frac{\mathcal N_G}{\mathcal N}}\frac{\varepsilon_G^2}{\tL^2} \right)
\\
&+
\frac{\mathcal N}{2}
\left[
\ln\left(
\frac{\widetilde L^{\,2}+m^2}
{\varepsilon^2+m^2}
\right)
-\frac12
+\frac{m^2}{\widetilde L^{\,2}}
-\frac{m^4}{\widetilde L^{\,4}}
\ln\left(
\frac{\widetilde L^{\,2}+m^2}{m^2}
\right)
\right].
\end{aligned}
\label{one-loopG}
\end{equation}

\subsection{One loop Equation of motion}

As in the classical case we work at fixed $\mathcal N$ first and
vary the gravitational scale $\widetilde L$ at fixed $\mathcal N$ to obtain the cardinal RG=GR equation $\left.\partial_{\widetilde L}\Gamma\right|_{\mathcal N}=0$.
To evaluate the stationary points it is convenient to use that $\tL^{2}
\frac{\pa \Box_g}
{\partial\tL^{\,2}} = \Box_g$, and therefore obtain that  for the fixed-cardinality trace,
\begin{align}
\tL^{2}
\frac{\partial\Gamma_1}
{\partial\tL^{\,2}}
&=
\frac12\Tr_{\mathcal N}
\tL^{2}
\frac{\partial }
{\partial\tL^{\,2}}
\ln\left(\frac{-\Box_g+m^2}{\mu^2}\right)
\nonumber = \frac12\Tr_{\mathcal N}
\left(\frac{\Box_g}{\Box_g+m^2}\right)
\nonumber\\
&=\frac12\Tr_{\mathcal N}\mathbb I
- \frac12\Tr_{\mathcal N}
\left(\frac{m^2}{\Box_g+m^2}\right).
\end{align}
Therefore
\begin{align}
\tL^{2}
\frac{\partial\Gamma_1}
{\partial\tL^{\,2}}
=
\frac{\mathcal N}{2}
\left[
1-\left.
{2 m^2}
\frac{\partial\gamma_1}{\partial m^2}
\right|_{\mu}
\right],
\end{align}
which means that we have the  bound
\begin{equation}
\left.
{2m^2}
\frac{\partial\gamma_1}{\partial m^2}
\right|_{\mu}
 =  \frac1{\mathcal N}\Tr_{\mathcal N}
\left(
\frac{m^2}{ \Box_g+ m^2}
\right) \quad \Rightarrow \quad
0
\leq
\left.
{2m^2}
\frac{\partial\gamma_1}{\partial m^2}
\right|_{\mu}
\leq1.
\label{CardinalityEquality:102}
\end{equation}
Moreover we see that
the maximum of $m^2 \pa_{m^2}\gamma_1$ is attained in the limit of infinite mass and the minimum is reached at zero mass.
We use the trace identity \eqref{traceid} to obtain the inequality.
The stationarity condition therefore reads
\begin{equation}
\frac{\varepsilon_G^4}{\tL^4}
-
\sqrt{\frac{\mathcal N_G}{\mathcal N}} \frac{\varepsilon_G^2}{\tL^2}
=
\frac14 \left[
1-\left.
{2 m^2}
\frac{\partial\gamma_1}{\partial m^2}
\right|_{\mu}
\right].
\label{CardinalityEquality:100}
\end{equation}
In the following we denote by $\tL_*^2(m, \mathcal N)$ the positive root of this equation. We show below that the solution always exists and is unique.
 The condition \eqref{CardinalityEquality:102} means that  the RHS of this equation  is positive and smaller than $\frac14$. Using \eqref{Gammaex}, it can be explicitly evaluated
\begin{equation}
\left.
{m^2}
\frac{\partial\gamma_1}{\partial m^2}
\right|_{\mu}
=
\frac{m^2}{\tL^{\,2}}
\left[
1-
\frac{m^2}{\tL^{\,2}}
\ln\left(
1+\frac{\tL^{\,2}}{m^2}
\right)
\right].
\label{CardinalityEquality:101}
\end{equation}

\subsection{Analyzing the one-loop equation of motion}
The previous analysis  establishes that the one loop equation of motion  with respect to $\tL$ is
\begin{equation}
\frac{\varepsilon_G^4}{\tL^4}
-
\sqrt{\frac{\mathcal N_G}{\mathcal N}} \frac{\varepsilon_G^2}{\tL^2}
= \frac14 \tau\left(\frac{\tL^2}{m^2}\right)
\label{eom}
\end{equation}
where the function on the RHS is
\be
\tau(x)= 1-\frac2{x}\left(1- \frac{\ln(1+x)}{x}\right) = \frac1{x^2}\int_0^x\frac{2t^2}{1+t}\,\dd t
\ee
It is a function that satisfies, for $x\geq 0$, the conditions
\be\label{taucond}
0\leq \tau(x)\leq 1,\qquad \tau'(x)>0, \qquad \tau+\frac{x}{2}\tau' =\frac{x}{1+x}.
\ee
$\tau$ represents the 
expectation value of the total stress per degree of freedom. The useful endpoint expansions are
\begin{align}
 \tau(x)&=\frac23x-\frac12x^2+\frac25x^3-\frac13x^4+O(x^5),
 &&x\ll1,
\label{eq:Csmall}\\
 \tau(x)&=1-\frac2x+\frac{2\ln x}{x^2}+O(x^{-3}),
 &&x\gg1.
\end{align}
To analyze the solution of \eqref{eom} it is convenient to introduce dimensionless variables that represent the momentum scale and mass in units of the vacuum energy scale.
\begin{align}\label{defsymb}
\stl:=\frac{\tL^2}{\varepsilon_G^2},
 \qquad
\sm:=\frac{m^2}{\varepsilon_G^2}, \qquad \sr:= \sqrt{\frac{\mathcal N}{\mathcal N_G}}.
\end{align}
We also introduce the ratio $\sr$, which is smaller than $1$ when $\mathcal N<\mathcal N_G$.
Equation~\eqref{eom} can be written as the condition $\mathcal E=0$, where
\begin{equation}
\cE(\stl;\sr,\sm):=\frac{\stl}{\sr}+\frac{\stl^2}{4}\tau\left(\frac{\stl}{\sm}\right)-1.
\label{eq:Qdef}
\end{equation}
A central simplification comes from showing that $\cE$ is a strictly increasing and convex function of $\stl$
\begin{equation}
\pa_{\stl} \cE=\frac{1}{\sr}+\frac{\stl^2}{2(\sm+\stl)}>0,\qquad
\pa^2_{\stl} \cE=\frac{\stl(2\sm+\stl)}{2(\sm+\stl)^2}>0.
\label{eq:D}
\end{equation}
Since $\cE(0)=-1$ and $\cE(\stl)\to+\infty$ as $\stl\to\infty$, there is precisely one positive root denoted $\stl_\star(\sm,\sr)$. The dimensionful root is given by
\be\label{dimroot}
\tL^2_*(\mathcal N,m)=\varepsilon_G^2\stl_\star\left(\frac{m^2}{\varepsilon_G^2},\sqrt{\frac{\mathcal N}{\mathcal N_G}}\right)
\ee
Importantly, the function $\stl_\star$ increases with both $\sm$ and $\sr$. If we denote  $\cE'_*=\pa_{\stl}\cE(\stl_\star;\sr,\sm)$. Implicit differentiation gives
\begin{equation}
\pa_{\sr}\stl_\star =\frac{\stl_\star}{\sr^2 \cE'_*}>0,\qquad
\pa_{\sm}\stl_\star = \frac{\stl_\star^3 }{4 \sm^2 \cE'_*}  \tau'\left(\frac{\stl_\star}{\sm} \right) >0,
\label{eq:responses}
\end{equation}
From this we conclude that, at fixed $\mathcal N$, $\tL_*^2$ is bounded from above by its infinite-mass limit and from below by its zero-mass value. This means that
\be \label{boundtL}
\frac{2 \sr \varepsilon_G^2}{1+\sqrt{1+\sr^2}} \leq \tL^2_*(m, \mathcal N) \leq \sr \varepsilon_G^2
\ee

The inequalities~\eqref{boundtL} can also be translated into bounds on the
spacetime radius. Using Eq.~\eqref{RtR}, we have
\begin{equation}
L_*^2=\frac{\sqrt{12\mathcal N}\,\hbar^2}{\tL_*^2}
=\ell_c^2\frac{\sr\varepsilon_G^2}{\tL_*^2},
\qquad
\ell_c^2=\frac{\sqrt{12\mathcal N_G}\,\hbar^2}{\varepsilon_G^2}.
\end{equation}
Therefore,
\be \label{boundL}
\ell_c^2\leq L_*^2(m,\mathcal N)
\leq \frac{1+\sqrt{1+\sr^2}}{2}\,\ell_c^2.
\ee
The lower endpoint is attained for $m=0$, while the upper endpoint is approached for $m\to\infty$.

The stationary solution determines the effective
cosmological constant, defined by the curvature of the
quantum-corrected de Sitter geometry,
$\Lambda_{\rm eff}=3/L_*^2$. Using the fixed-cardinality
relation and the definitions above, we obtain
\begin{equation}
\boxed{
\Lambda_{\rm eff}(\mathcal N,m)
=\Lambda\,\frac{\stl_\star(\sm,\sr)}{\sr},
\qquad
\sr=\sqrt{\frac{\mathcal N}{\mathcal N_G}}.
}
\label{eq:Lambdaeff_stationary}
\end{equation}
It is interesting to consider the large-cardinality limit
at fixed gravitational parameters and scalar mass.
As $\mathcal N/\mathcal N_G\to\infty$, the stationary
equation implies that $\stl_\star$ approaches a finite value
$\stl_\infty(\sm)$, determined by
\begin{equation}\label{tildelinf}
\frac{\stl_\infty^2}{4}
\tau\left(\frac{\stl_\infty}{\sm}\right)=1.
\end{equation}
Consequently,
\begin{equation}
\Lambda_{\rm eff}(\mathcal N,m)
\simeq
\Lambda\,\stl_\infty(\sm)
\sqrt{\frac{\mathcal N_G}{\mathcal N}},
\qquad
\mathcal N\gg\mathcal N_G.
\label{eq:Lambdaeff_largeN}
\end{equation}

Thus, at fixed $\Lambda$, $G$ and scalar mass, the
effective cosmological constant decreases as
$\mathcal N^{-1/2}$ for large cardinality. The
corresponding de Sitter radius grows as
$\mathcal N^{1/4}$, and the stationary geometry
approaches flat space.

The function $\stl_\infty(\sm)$ is the unique  solution of the equation
\be
\int_0^{\stl_\infty}\frac{2t^2}{\sm+t}\,\dd t=4.
\ee
As such it can be shown to be an increasing function of $\sm$ and is given in the limiting behavior $m\to 0$ and $m\to \infty$ by (see Fig.~\ref{fig:c-range} below)
\begin{align}
\stl_\infty&= 2+\sm+\sm^2\left[\frac14-\frac12\ln\left(\frac2\sm\right)\right]+O(\sm^3|\ln\sm|),\\
\stl_\infty &=(6\sm)^{1/3}+\frac{3}{2 (6\sm)^{1/3}}+O(\sm^{-1}).
\end{align}

These results agree with those obtained by Becker and
Reuter~\cite{Becker:2020mjl,Becker:2021pwo} and
Ferrero and Percacci~\cite{Ferrero:2024yvw}.
For a massless scalar we recover exactly their
self-consistent de Sitter solution, identifying
our cardinality $\mathcal N$ with the number of
retained modes $f(K)$ in their approach.
For a massive scalar, our construction agrees with the
leading large-cutoff curvature equation obtained
by Ferrero and Percacci \cite{Ferrero:2024yvw}.\footnote{
In the spectral approach of \cite{Becker:2020mjl,Becker:2021pwo}, \cite{Ferrero:2024yvw}, the cutoff
$K$ restricts the eigenmodes of the Laplacian,
while in our case the cardinality $\mathcal N$
is the volume of the accessible
phase-space region. The restricted trace is then
evaluated at leading Weyl order.
For a massless scalar all retained modes contribute
equally to the effective-action stress tensor, and
the result depends only on their number $f(K)$.
For a massive scalar each mode contributes with
a weight $E_n/(E_n+m^2)$, where  $E_n$ denotes the eigenvalue of
the positive scalar Laplacian on the four-sphere $S^4_L$, and the result depends
on the detailed distribution of the eigenvalues.
Our momentum-space integral corresponds to the
leading Weyl approximation to this spectral sum.
For this reason the massless results coincide
exactly, while the massive results agree at
leading order in the large-cardinality expansion.
}

In the next section we
use the thermodynamic entropy of the coupled
gravity--matter system and impose a holographic
entropy bound. This will provide an additional
relation between the cardinality and the
normalization scale of the relevant quantum theory.

\subsection{QFT limit}

An important connection with conventional QFT emerges by considering a
correlated limit in which gravity decouples and the background becomes flat,
while the momentum scale governing the matter loop cutoff stays finite. This can be achieved by taking the \emph{double scaling  limit}
$G,\Lambda\to0$ that keeps  ratios $\Lambda/G$ and $\mathcal{N}/\mathcal{N}_G$ fixed.
More precisely, we consider the flat non-gravitational  limit
\be
\cN_G= \frac{3\pi}{\hbar G\Lambda} \to \infty, \qquad \varepsilon_G^2=\sqrt{\frac{4\pi\hbar^3\Lambda}{G}}\to \varrho^2 \quad \text{and}\quad \sr \to r\,\,\text{arbitrary}.
\ee
Keeping $\varrho>0$ and $r>0$ finite entails
$\Lambda/G\to\varrho^4/(4\pi\hbar^3)$ and
$\cN_G\sim12\pi^2\hbar^2/(G^2\varrho^4)\to\infty$.
Moreover, $\sr=\sqrt{\cN/\cN_G}\to r$ means that
$\cN\sim r^2\cN_G$: both the gravitational cardinality and the
accessible phase-space cardinality grow, while their ratio remains finite.
The stationary momentum scale nevertheless approaches the finite value
\begin{align}
\tL_*^2\longrightarrow
\varrho^2\stl_\star\!\left(\frac{m^2}{\varrho^2},r\right),
\end{align}
as follows from Eq.~\eqref{dimroot}. At the same time,
$\ell_c^2=3/\Lambda\to\infty$, and the fixed-cardinality relation
implies $L_*^2/\ell_c^2\to r/\stl_\star$; hence the scalar curvature
$R=12/L_*^2$ vanishes. Thus this limit yields a flat background and a
finite range of momentum integration. The bounds~\eqref{boundtL} show that
$\tL_*$ remains of order $\varrho$ for any fixed positive $r$;
in particular $\tL_*^2\to r\varrho^2$ for $m\gg\varrho$.

This provides a concrete bridge to the standard QFT calculation of
Sec.~III. At leading Weyl order, the matter contribution per unit
spacetime volume becomes
\begin{align}
\frac{\hbar^4\Gamma_1}{\mathcal V}
\longrightarrow
\frac12\int_{|p|\leq\tL_*}\frac{d^4p}{(2\pi)^4}
\ln\!\left(\frac{p^2+m^2}{m^2+\varepsilon^2}\right),
\end{align}
which is precisely the conventional cutoff integral of
Eq.~\eqref{eq:V1_cutoff}, with the normalization prescription used here.
The loop integral is therefore the same in the two descriptions.
The difference is how its upper momentum scale is treated: in conventional
cutoff QFT the cutoff scale is  prescribed independently of any physical parameters and can be chosen to be arbitrarily large.  In our context this scale is fixed by the parameter $\rho$ which emerges from the expression for the vacuum energy in the flat space $\Lambda \to 0$
and non gravitational limit $G \to 0$,  that is for scales much smaller than
the cosmological scale and much larger than the
Planck scale.

Moreover, the relation $\varrho^4=4\pi\hbar^3\Lambda/G$ explains why $\varrho$
has the scaling of a QFT ultraviolet scale. If one estimates the
cosmological term from a quartically cutoff-sensitive vacuum contribution,
$\Lambda/G\propto\Lambda_{\rm UV}^4$, then $\varrho$ is of the same order as
that cutoff, up to coefficients and matching conventions.
This shows that in the flat non-gravitational
limit $\varrho$ sets the scale of the effective
momentum cutoff, whose precise value also depends
on the mass and the cardinality ratio.

The precise result established here is that a finite flat-space matter momentum scale
survives the gravitational decoupling limit of the cardinal framework. This suggests that quantum field theory should be defined, in terms of gravitational (and in general, quantum gravitational) data, in a fully covariant and background independent manner.

\section{Entropy bound}\label{Sect:EntropyBound}

The field equation derived above has as a solution
a stationary momentum scale
$\widetilde L_\star$ for every positive cardinality
$\mathcal N$, but  $\mathcal N$
was undetermined. To proceed, in this section we consider the entropy
of the coupled gravity--matter system and
a holographic entropy bound.  As we will see, the
resulting condition can be used to  fix $\varepsilon$ as a function of
$\mathcal N$.

\subsection{Entropy of the coupled system}

We first recall the Gibbons--Hawking construction
\cite{Gibbons:1976ue}, \cite{Gibbons:1977mu}. The Euclidean continuation of de Sitter space is
a four-sphere $S^4_L$. Regularity at the cosmological horizon fixes the inverse
temperature to be  $\beta=2\pi L/\hbar$.
The Euclidean gravitational path integral gives the partition function,
which is dominated in the semiclassical approximation by the regular
stationary solution,
\begin{equation}
 Z\simeq \exp(-\Gamma_0|_\star).
 \label{eq:VI_classical_Z}
\end{equation}
The thermodynamic relation between the entropy and the partition
function is
\begin{equation}
 S=\beta E+\log Z,\qquad
 E=-\frac{\partial\log Z}{\partial\beta}.
 \label{eq:VI_thermo_identity}
\end{equation}
For the compact de Sitter solution the gravitational Hamiltonian is a
constraint, and there is no asymptotic boundary term that could give a
nonzero total energy. In the Gibbons--Hawking ensemble one therefore has
$E=0$ and obtains
\begin{equation}
 S_{\rm GH}=-\Gamma_0|_\star
 =\frac{\pi L_c^2}{\hbar G}
 =\frac{3\pi}{\hbar G\Lambda}=\mathcal N_G,
 \qquad L_c^2=\frac{3}{\Lambda}.
 \label{eq:VI_classical_entropy}
\end{equation}

The same construction can be applied to the total one loop effective action. We still have a gravitational path integral on a compact
Euclidean geometry; the difference is that the exponent now contains the
effective action $\Gamma=\Gamma_0+\Gamma_1$. The stationary
geometry is determined by the full action, with the restriction to the
accessible phase-space region imposed throughout. Therefore we write
\begin{equation}
 Z_{\mathcal N}(\beta;m,\mu)
 \simeq\exp\bigl[-\Gamma
 (\widetilde L_\star,\mathcal N;m,\mu)\bigr].
 \label{eq:VI_partition_tot}
\end{equation}
Here $\mathcal N$ is kept fixed when the stationary equation is solved,
and $m$ and the determinant normalization $\mu$ are held fixed in the
variation. In particular, the matter one-loop contribution is part of the
thermodynamic partition function; it is not treated as an external source
whose contribution to entropy is added afterward.

It is useful to check the energy term explicitly in the one-parameter
family considered here. Since $\beta\propto L$, while fixed
cardinality gives $L^4\widetilde L^4=12\mathcal N\hbar^4$, a variation of
the inverse temperature along this family satisfies
\begin{equation}
 \left.\beta\frac{\partial}{\partial\beta}\right|_{\mathcal N,m,\mu}
 =\left.L\frac{\partial}{\partial L}\right|_{\mathcal N,m,\mu}
 =-2\left.\stl
   \frac{\partial}{\partial\stl}\right|_{\mathcal N,m,\mu},
 \qquad
 \stl:=\frac{\widetilde L^2}{\varepsilon_G^2}.
 \label{eq:VI_beta_variation}
\end{equation}
The stationary equation of Sec.~V then implies
\begin{equation}
 \left.\beta\frac{\partial\Gamma_{\rm tot}}{\partial\beta}
 \right|_{\mathcal N,m,\mu;\star}
 =-2\stl_\star
 \left.\frac{\partial\Gamma_{\rm tot}}
 {\partial\stl}\right|_{\mathcal N,m,\mu;\star}=0.
 \label{eq:VI_energy_zero}
\end{equation}
Thus the vanishing of the thermodynamic energy has a direct expression
in the fixed-cardinality model. It is consistent with the Hamiltonian
constraint argument: the total action is stationary with respect to the
scale which also determines the Euclidean temperature. With this
thermodynamic prescription we identify the entropy of the coupled
system with
\begin{equation}
 S(\mathcal N,m):=-\Gamma
 \bigl(\widetilde L_\star(\mathcal N,m),\mathcal N;m,\mu\bigr).
 \label{eq:VI_entropy_definition}
\end{equation}
In the absence of the one-loop matter correction this reduces to
Eq.~\eqref{eq:VI_classical_entropy}. Equation~\eqref{eq:VI_entropy_definition}
is our extension of the Gibbons--Hawking prescription to the
cardinality-restricted one-loop effective action.

The entropy in Eq.~\eqref{eq:VI_entropy_definition} is assigned to a
stationary configuration at fixed $\mathcal N$. We next compare the
entropies of the stationary solutions obtained for different values of
$\mathcal N$. This second step is an entropy-selection prescription; it
does not change the condition that the field equation be solved at
fixed cardinality.

\subsection{Holographic bound}

While the generalized second law \cite{bekenstein1974generalized, wall2009ten, wall2012proof} established that the sum of gravitational entropy plus external matter entropy always increase in time,
the holographic principle~\cite{tHooft:1993dmi, Susskind:1994vu, Bousso:2002ju} suggests that the entropy of a gravitational
system cannot be arbitrarily large compared with the area of an
appropriate surface. Here we propose that the Gibbons--Hawking area entropy
continues to provide an upper bound on the thermodynamic
entropy of the coupled gravity--matter system when
one-loop corrections are included. The holographic bound therefore reads
\begin{equation}
 S(\mathcal N,m)\leq S_{\rm GH}(L_\star),
 \qquad
 S_{\rm GH}(L_\star)
 =\frac{A_H(L_\star)}{4\hbar G}
 =\frac{\pi L_\star^2}{\hbar G}.
 \label{eq:VI_bound}
\end{equation}
Here $L_\star$ is determined by the one-loop field
equation and depends on $\mathcal N$ and $m$.

Our proposal is closely related to Bousso's $D$-bound
\cite{Bousso:2000md}. The latter considers a matter
system inside the cosmological horizon of an asymptotically
de Sitter spacetime. When the matter crosses the horizon,
the geometry approaches empty de Sitter space with the
same cosmological constant $\Lambda$. The generalized
second law then requires
\begin{equation}
 S_{\rm matter}+\frac{A_c}{4\hbar G}
 \leq \frac{A_0}{4\hbar G}
 =\mathcal N_G,
 \label{eq:VI_Dbound}
\end{equation}
where $A_c$ is the original cosmological horizon area
and $A_0$ is the horizon area of empty de Sitter space.

If we identify the total entropy on the left-hand side
of the $D$-bound with our thermodynamic entropy $S$,
the difference between the two bounds is in the
right-hand side. Bousso's bound uses the entropy
$\mathcal N_G$ of empty de Sitter space, whereas
ours uses the horizon entropy of the quantum-corrected
stationary geometry. Since the solutions of Sec.~V
satisfy $L_\star^2\geq\ell_c^2$, we have
$S_{\rm GH}(L_\star)\geq\mathcal N_G$.
Our proposed bound is therefore weaker than the
corresponding $D$-bound inequality.

\subsection{Entropy maximization and the field equation}

To make the argument explicit, we use the variables of Sec.~V, and normalization of the
one-loop determinant is chosen as in Sec.~II
\begin{equation}
 \mu^2=m^2+\varepsilon^2,
 \qquad \varepsilon>0,
 \label{eq:VI_mu_def}
\end{equation}
and $\mu$ is held fixed when the field equation is varied. With the
function $\tau$ introduced in Sec.~V, the stationary equation is
\begin{equation}
 \frac{\stl_\star}{\sr}
 +\frac{\stl_\star^2}{4}
 \tau\!\left(\frac{\stl_\star}{\sm}\right)=1,
 \qquad
 \tau(x)=1-\frac{2}{x}\left(1-\frac{\log(1+x)}{x}\right).
 \label{eq:VI_stationary}
\end{equation}
where $\stl_\star=\stl_\star(\mathcal{N})$ is a function of $\mathcal{N}$.
The function $\tau$ increases from zero to one and is positive for
positive argument. The massless limit corresponds to $\tau=1$.

We now evaluate the entropy on these stationary solutions. Using the cardinality relation \eqref{RtR}, and the vacuum energy relationship
$
\sqrt{\mathcal{N_G} }\varepsilon_G^2 = \frac{\sqrt{12} \pi \hbar }{G}
$, we evaluate the
Gibbons--Hawking entropy to be
\begin{equation}
 S_{\rm GH}(L_\star)
 = \frac{\pi L_\star^2}{\hbar G}
 = \sqrt{\mathcal N\mathcal N_G} \frac{\varepsilon_G^2}
 {\tL^2_\star}
 =\mathcal N_G\frac{\sr}{\stl_\star}.
 \label{eq:VI_area_entropy}
\end{equation}
Combining Eqs.~\eqref{eq:Lambdaeff_stationary} and \eqref{eq:VI_area_entropy}, we obtain the
relation
\begin{equation}
\boxed{
\Lambda_{\rm eff} S_{\rm GH}(L_\star)
=\Lambda\mathcal N_G
=\frac{3\pi}{\hbar G}.
}
\end{equation}
This relation is one of the most important results of this paper.

Let us turn to the dependence of these
quantities on cardinality. From Eqs.~\eqref{tildelinf} and \eqref{eq:VI_area_entropy},
for fixed gravitational parameters and scalar mass,
we find
\begin{equation}
S_{\rm GH}(L_\star)
\simeq
\frac{\sqrt{\mathcal N\mathcal N_G}}
{\stl_\infty(\sm)},
\qquad \mathcal N\gg\mathcal N_G.
\end{equation}
Thus the horizon entropy grows as $\sqrt{\mathcal N}$,
while the effective cosmological constant decreases
as $1/\sqrt{\mathcal N}$. We see that in the large $\mathcal N$ limit the effective cosmological constant $\Lambda_{eff}$ is inverse proportional to the entropy, confirming the findings of \cite{Freidel:2022ryr}, \cite{Freidel:2023ytq}.

Using Eq.~\eqref{eq:VI_stationary} in the full effective action \eqref{one-loopG}, one
finds a particularly simple expression for its on-shell value. Indeed,
for the reduced one-loop action $\gamma_1$ of Sec.~II one has
\begin{equation}
 \gamma_1(\stl)
 +\frac14\tau\!\left(\frac{\stl}{\sm}\right)
 =\frac12\log\frac{m^2+\varepsilon_G^2\stl}
 {m^2+\varepsilon^2}.
 \label{eq:VI_gamma_identity}
\end{equation}
Together with the gravitational part of the action this gives
\begin{equation}
 S(\mathcal N,m)
 =S_{\rm GH}(L_\star)
 -\frac{\mathcal N}{2}
 \log\frac{m^2+\widetilde L_\star^2}{m^2+\varepsilon^2}.
 \label{eq:VI_on_shell_entropy}
\end{equation}
Equation~\eqref{eq:VI_on_shell_entropy} shows that the difference
between the total on-shell entropy and the Gibbons--Hawking
area entropy depends on the normalization scale $\mu$ of the
one-loop determinant and on the stationary momentum scale.
The holographic bound therefore restricts their relative values. The bound~\eqref{eq:VI_bound} reduces to
\begin{equation}
 \widetilde L_\star^2\geq\varepsilon^2.
 \label{eq:VI_bound_l}
\end{equation}
Thus the holographic condition restricts the stationary solutions,
without requiring an upper bound on their cardinality.

We next maximize $S$ with respect to $\mathcal N$, treating $m$ and
$\varepsilon$ as fixed parameters. The stationary value
$\stl_\star$ depends on $\mathcal N$, but its
derivative does not contribute to the first derivative of the on-shell
action, since the full action is stationary with respect to
$\stl$. It follows that
\begin{equation}
 \frac{dS}{d\mathcal N}
 =-\left.\frac{\partial\Gamma_{\rm tot}}{\partial\mathcal N}
 \right|_{\stl=\stl_\star}
 =-\frac12\log\frac
 {m^2+\tL^2_\star(\mathcal{N})}{m^2+\varepsilon^2}.
 \label{eq:VI_first_derivative}
\end{equation}
Equations~\eqref{eq:VI_on_shell_entropy} and
\eqref{eq:VI_first_derivative} also imply the identity
\begin{equation}
 S_{\rm GH}(L_\star)
 =S-\mathcal N\frac{dS}{d\mathcal N}.
 \label{eq:VI_entropy_derivative_identity}
\end{equation}
The entropy bound is therefore equivalent to
$dS/d\mathcal N\leq0$ hence $L_\star^2(\mathcal{N})\geq\varepsilon^2$. The allowed stationary solutions lie on the
decreasing side of the entropy curve.

As shown in Sec.~V, $\stl_\star$ increases
monotonically with $\mathcal N$. Consequently,
\begin{equation}
 \frac{d^2S}{d\mathcal N^2}
 =-\frac{\varepsilon_G^2}
 {2(m^2+\varepsilon_G^2\stl_\star)}
 \frac{d\stl_\star}{d\mathcal N}<0.
 \label{eq:VI_concavity}
\end{equation}


The entropy is therefore strictly concave. The
stationary field equation determines a unique solution
$\stl_\star(\mathcal N,\sm)$ for each positive cardinality,
 while the entropy maximization uniquely relates the value of the reference  scale $\varepsilon$ in terms of $\stl_\star$. This means that after entropy maximization both the momentum scale $\tL$ and the reference scale are determined by $\mathcal{N}$ and $m$.
 At this value the equilibrium entropy is a function of $\mathcal{N}$, $S_{\text{eq}}(\mathcal{N})= S_{\text{GH}}(L_*(\mathcal{N}) $.

The equilibrium occurs at the value
\begin{equation}
\varepsilon_{\rm eq}^2={\varepsilon_G^2}  \stl_\star(\mathcal N,\sm),
 \qquad
 S_{\max}=S_{\rm GH}(L_{\rm eq}).
\end{equation}
For
fixed $m$ and $\varepsilon$, its maximum is unique
and is reached when
$\widetilde L_\star^2=\varepsilon^2$.
We can also read this relation in the opposite
direction. Taking the cardinality $\mathcal N$
as the physical parameter, the entropy condition
determines the normalization scale $\varepsilon$
in terms of $\mathcal N$ and $m$. The corresponding
equilibrium entropy is then a function of
cardinality, as shown in Fig.~1.

\begin{figure}[h!]
\centering
\includegraphics[width=0.85\textwidth]{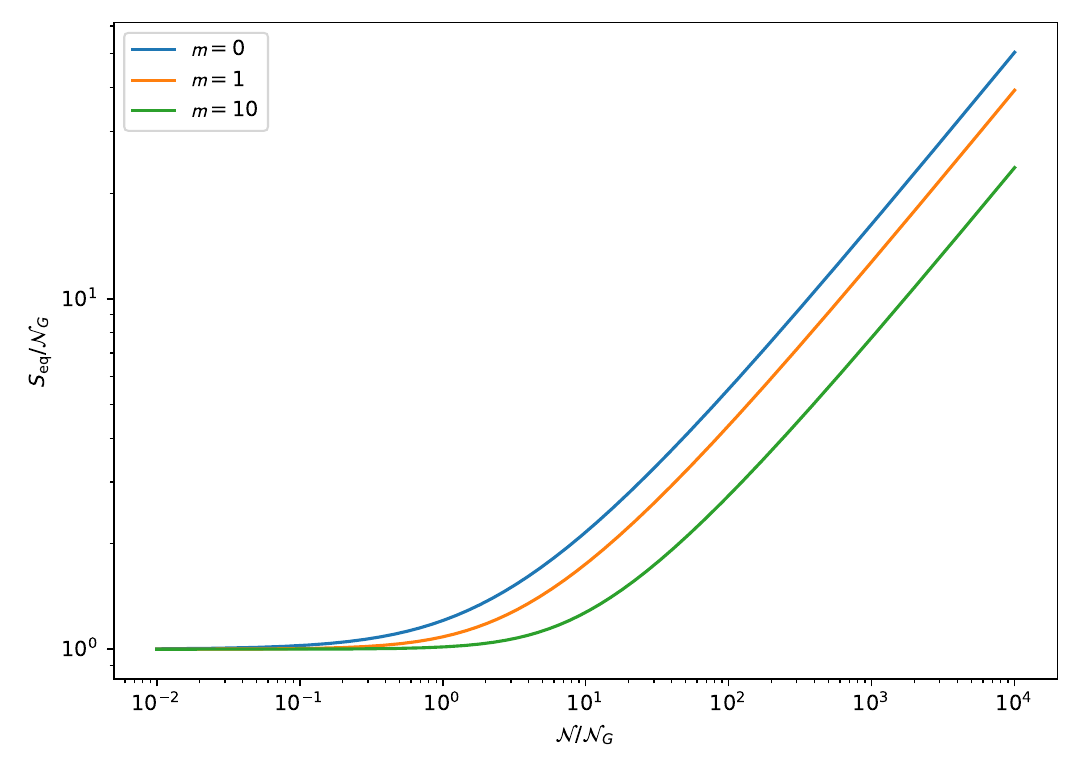}
\caption{The equilibrium entropy as a function of cardinality for different values of $m$.}
\label{fig:S(L_*}
\end{figure}

\begin{figure}[h!]
\centering
\includegraphics[width=0.85\textwidth]{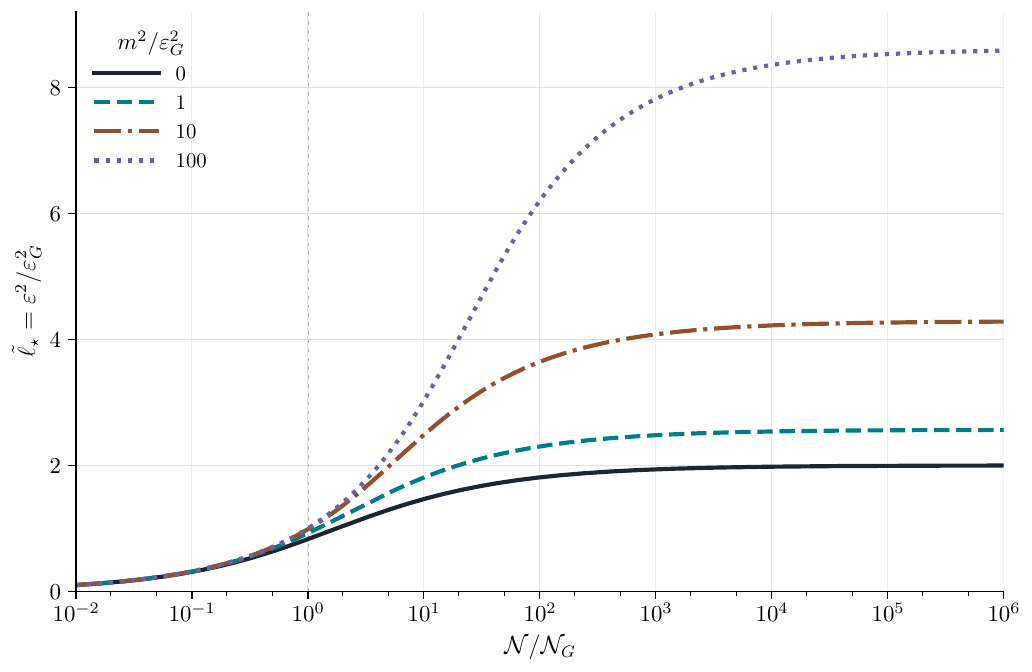}
\caption{The function $\varepsilon^2/\varepsilon^2_G = \stl_{\star}(\mathcal{N}/\mathcal{N}_G,m^2/\varepsilon^2_G)$ for different values of $m$.}
\label{fig:e(N)}
\end{figure}

\begin{figure}[h!]
\centering
\includegraphics[width=0.85\textwidth]{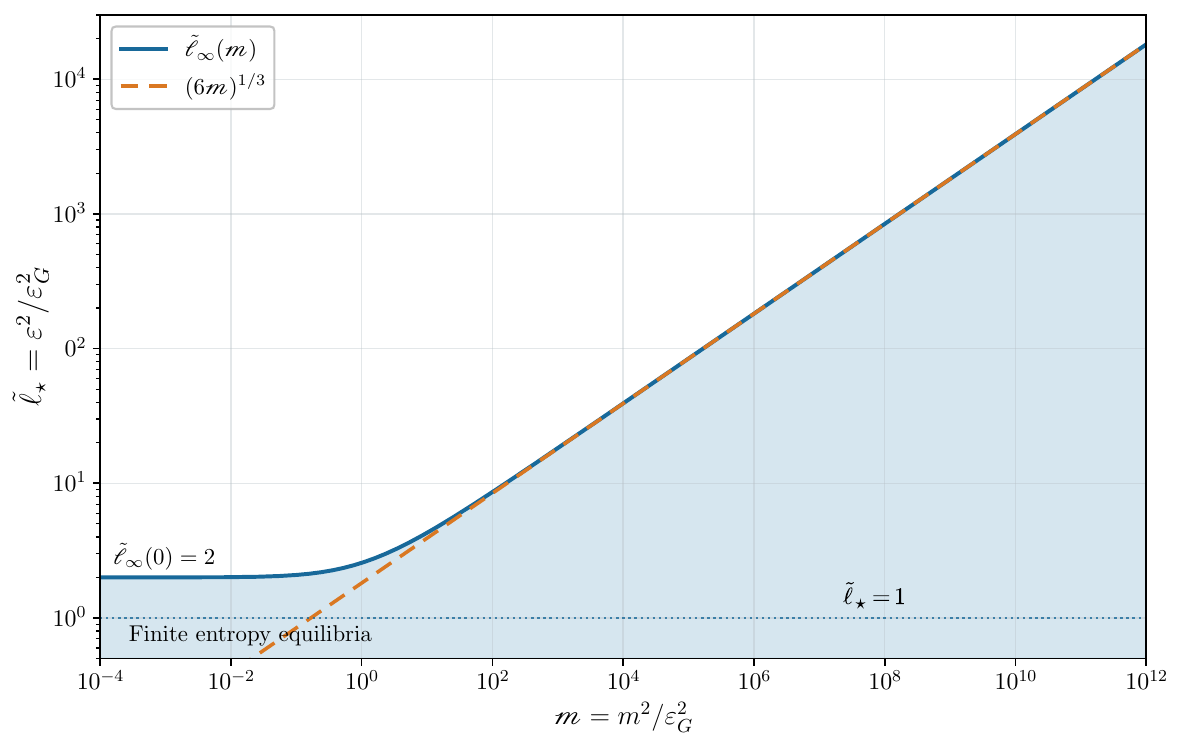}
\caption{Admissible values of the equilibrium normalization parameter $\stl_\star=\varepsilon^2/\varepsilon_G^2$ as a function of the dimensionless scalar mass $\sm=m^2/\varepsilon_G^2$. The shaded region corresponds to finite entropy equilibria, $0<\stl_\star<\stl_{\infty}$. The dashed curve shows the large-mass approximation $\stl_{\infty}\simeq(6\sm)^{1/3}$.}
\label{fig:c-range}
\end{figure}

Thus the entropy condition determines the reference
scale $\varepsilon$ in terms of the cardinality
$\mathcal N$ and the scalar mass $m$.
Substituting $\stl_\star=\varepsilon^2/\varepsilon_G^2$
into the stationary field equation gives
\begin{equation}
 \sr=\frac{\stl_\star}{D(\stl_\star,\sm)},
 \qquad
 D(\stl,\sm)=1-\frac{\stl^2}{4}
 \tau\left(\frac{\stl}{\sm}\right),
 \label{eq:entropy_cardinality}
\end{equation}
where $\sr=\sqrt{\mathcal N/\mathcal N_G}$.

The function $D$ decreases monotonically,
\begin{equation}
 \frac{\partial D}{\partial\stl}
 =-\frac{\stl^2}{2(\sm+\stl)}<0,
 \qquad D(0,\sm)=1.
\end{equation}
It has a unique positive zero $\stl_\infty(\sm)$,
 defined in Eq.~(74). Therefore, the
relation between $\sr$ and $\stl_\star$ is
one-to-one, with
\begin{equation}
 \boxed{
 0<\stl_\star<\stl_\infty(\sm),
 \qquad
 0<\varepsilon^2<
 \varepsilon_G^2\stl_\infty(\sm).
 }
\end{equation}
As $\stl_\star$ approaches $\stl_\infty$,
the corresponding cardinality diverges.
For the massless scalar $\stl_\infty=2$, while
for large mass its behavior is given by Eq.~(78).
The dependence of $\stl_\star$ on cardinality
is shown in Fig.~\ref{fig:e(N)} and the admissible values of $\stl_\star$ are depicted in Fig.~\ref{fig:c-range}.

The equilibrium condition also has a flat-space
limit. In the double-scaling limit of Sec.~V.C,
where $\varepsilon_G^2\to\varrho^2$ and $\sr$
remains fixed, we obtain
\begin{equation}
 \frac{\varepsilon^2}{\sr\varrho^2}
 +\frac{\varepsilon^4}{4\varrho^4}
 \tau\left(\frac{\varepsilon^2}{m^2}\right)=1.
\end{equation}
Thus the relation between the momentum scale and
the normalization of the determinant survives
the decoupling of gravity.

\section{Stability of the cosmological constant}
\label{Sect:CCstability}

We now return to the cosmological constant problem.
In conventional EFT, vacuum energy receives contributions
proportional to the fourth power of an ultraviolet scale,
or of order $m^4$ when a heavy field is integrated
out~\cite{Weinberg:1988cp,Padilla:2015aaa}.
In the cardinal approach, the relevant question is instead
how quantum fluctuations modify the curvature of the
self-consistent de Sitter geometry.

As shown in Sec.~\ref{Sect:EntropyBound}, entropy
maximization fixes the normalization scale according to
$\widetilde L_{\rm eq}^2=\varepsilon^2$.
Using Eq.~\eqref{eq:Lambdaeff_stationary},
we compute the effective cosmological constant, defined by the curvature
$\Lambda_{\rm eff}=3/L_{\rm eq}^2$, is
\begin{equation}
 \frac{\Lambda_{\rm eff}}{\Lambda}
 =\frac{\stl_\star}{\sr}
 =1-\frac{\stl_\star^2}{4}
 \tau\left(\frac{\stl_\star}{\sm}\right).
 \label{eq:VII_Lambda_ratio}
\end{equation}
Since $0<\stl_\star<\stl_\infty(\sm)$ and
$\tau(x)>0$ for $x>0$, it follows that
\begin{equation}
0<\Lambda_{\rm eff}<\Lambda.
 \label{eq:VII_universal_bound}
\end{equation}
Thus, the one-loop scalar contribution always decreases
the curvature, but its magnitude cannot exceed the
bare cosmological constant.

A stronger bound follows from the monotonicity of
the stationary solution with respect to the scalar
mass, established in Eq.~\eqref{eq:responses}.
Using the massless solution and the infinite-mass
limit of Eq.~\eqref{boundtL}, we find
\begin{equation}
 \frac{2}{1+\sqrt{1+\mathcal N/\mathcal N_G}}
 \leq
 \frac{\Lambda_{\rm eff}}{\Lambda}<1.
 \label{eq:VII_bound_fixed_cardinality}
\end{equation}
For $\mathcal N=\mathcal N_G$, the relative
correction is therefore bounded by
$3-2\sqrt{2}\simeq0.172$.
More generally, for a massless scalar
the effective cosmological constant is
\begin{equation}
 \Lambda_{\rm eff}(\mathcal N,0)
 =\frac{2\Lambda}
 {1+\sqrt{1+\mathcal N/\mathcal N_G}}.
 \label{eq:VII_massless}
\end{equation}

\begin{figure}[htbp]
 \centering
 \includegraphics[width=0.85\textwidth]
 {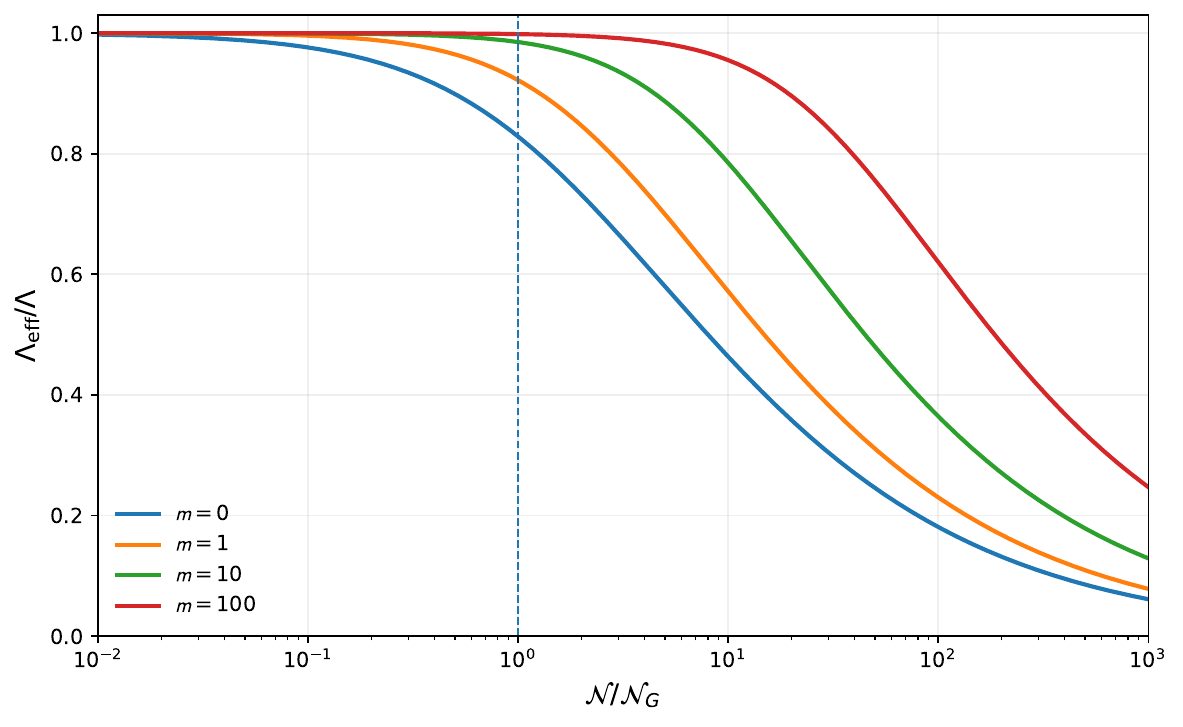}
 \caption{The effective cosmological constant
 $\Lambda_{\rm eff}/\Lambda$ as a function of
 $\mathcal N/\mathcal N_G$ for several values of
 $\sm=m^2/\varepsilon_G^2$.
 The dashed line denotes $\mathcal N=\mathcal N_G$.
 The gravitational parameters are held fixed.}
 \label{fig:VII_Lambda_cardinality}
\end{figure}

The large-cardinality behavior has already been
derived in Eq.~\eqref{eq:Lambdaeff_largeN}.
At fixed gravitational parameters and scalar mass,
$\Lambda_{\rm eff}\propto 1/\sqrt{\mathcal N}$,
and the quantum-corrected geometry approaches flat
spacetime. At the same time, the Gibbons--Hawking
entropy grows as $\sqrt{\mathcal N}$, as shown
in Sec.~\ref{Sect:EntropyBound}. Thus the effective
cosmological constant is inversely proportional
to the horizon entropy.

Finally, consider the heavy-mass limit at fixed
$\mathcal N/\mathcal N_G$ and fixed gravitational
parameters. Using the small-argument expansion
\eqref{eq:Csmall} in the equilibrium equation gives
$\stl_\star=\sr-\sr^4/(6\sm)+O(\sm^{-2})$.
Consequently,
\begin{equation}
 \frac{\Lambda_{\rm eff}-\Lambda}{\Lambda}
 =-\frac{\sr^3}{6\sm}+O(\sm^{-2}),
 \qquad \sm\to\infty.
 \label{eq:VII_heavy_fixed_cardinality}
\end{equation}
The correction therefore vanishes as $m^{-2}$.

These results prove the radiative stability
of the cosmological constant in the
cardinal one-loop model: the quantum correction is bounded
and heavy fields decouple at fixed cardinality.

\subsection{Large cardinality and the QFT limit}

The large-cardinality behavior depends on which
parameters are held fixed.

First, at fixed $G$, $\Lambda$, and physical mass $m$,
the gravitational cardinality $\mathcal N_G$ and
$\sm=m^2/\varepsilon_G^2$ remain constant.
As $\mathcal N\to\infty$, the equilibrium momentum
scale approaches the finite endpoint
$\stl_\infty(\sm)$ defined in
Eq.~\eqref{tildelinf}. Consequently,
\begin{equation}
 \Lambda_{\rm eff}\simeq
 \Lambda\,\stl_\infty
 \sqrt{\frac{\mathcal N_G}{\mathcal N}},
 \qquad
 L_{\rm eq}\simeq
 \frac{\ell_c}{\sqrt{\stl_\infty}}
 \left(\frac{\mathcal N}{\mathcal N_G}\right)^{1/4}.
 \label{eq:VII_largeN}
\end{equation}
The curvature thus vanishes as
$1/\sqrt{\mathcal N}$, while the
Gibbons--Hawking entropy grows as
$\sqrt{\mathcal N}$, which is consistent with
Eq.~\eqref{eq:VI_area_entropy}.

A different behavior arises if $G$ and
$\sr=\sqrt{\mathcal N/\mathcal N_G}$ are fixed,
while $\mathcal N,\mathcal N_G\to\infty$.
Using the definition of $\mathcal N_G$ and
Eq.~\eqref{eq:VII_Lambda_ratio}, we find
\begin{equation}
 \Lambda_{\rm eff}
 =\frac{3\pi}{\hbar G\mathcal N}
 \,\sr\,\stl_\star(\sr^2,\sm).
 \label{eq:VII_correlated}
\end{equation}
For fixed nonzero physical mass,
$\sm\to\infty$ and $\stl_\star\to\sr$, so that
$\Lambda_{\rm eff}\propto\mathcal N^{-1}$.
The same scaling holds for a massless scalar,
although with a different coefficient.

Of particular importance is the
\emph{double-scaling QFT limit}
introduced in Sec.~V.C:
\begin{equation}
 G,\Lambda\longrightarrow0,
 \qquad
 \frac{\Lambda}{G}=\mathrm{const},
 \qquad
 \frac{\mathcal N}{\mathcal N_G}=\mathrm{const}.
\end{equation}
In this limit, the gravitational cardinality
diverges, while the characteristic matter
momentum scale remains finite.
Indeed, $\varepsilon_G^2\to\varrho^2$ and
$\sm\to m^2/\varrho^2$.
The equilibrium equation therefore admits a
finite solution $\stl_\star$, and
\begin{equation}
 \widetilde L_{\rm eq}^2
 =\varepsilon_G^2 \stl_\star
 \longrightarrow \varrho^2 \stl_\star.
\end{equation}
On the other hand, the exact relation
\begin{equation}
 \boxed{
 \Lambda_{\rm eff}
 =\frac{\sqrt3}{2}
 \frac{\varepsilon_G^2}
 {\hbar^2\sqrt{\mathcal N}}\,
 \stl_\star(\sr^2,\sm)
 }
 \label{eq:VII_Lambda_vacuum_scale}
\end{equation}
shows that $\Lambda_{\rm eff}$ vanishes as
$1/\sqrt{\mathcal N}$.

Thus the double-scaling limit separates the
matter and gravitational scales: the curvature
vanishes, while the momentum domain entering
the one-loop determinant remains finite.
This provides the connection between the
cardinal gravitational framework and its
flat-spacetime QFT limit.

In the next section we apply the same prescription discussed in this section
to the stability of the Higgs mass.

\section{Stability of the Higgs mass}

We now turn to the stability of scalar masses.
In conventional EFT, scalar mass parameters are
sensitive to heavy thresholds, with corrections
of order the squared mass of a heavy particle \cite{Craig:2022eqo,Peskin:2025lsg}.
We investigate whether this sensitivity persists
when the one-loop effective potential is evaluated
on the entropy-selected gravitational background.

We consider two real scalars $\phi$ and $\psi$,
interacting through
\begin{equation}
 {\cal L}_{\rm int}
 =\frac{\lambda}{2}\phi^2\psi^2,
 \qquad \lambda>0,
\end{equation}
where $\psi$ has mass $m$, while $\phi$ represents
a Higgs-like scalar with classical mass parameter
$M_H$. Integrating out $\psi$ gives
\begin{equation}
 \Gamma_1[\phi]
 =\frac12\operatorname{Tr}_{\mathcal N}
 \ln\left(
 \frac{\Box_g+m^2+\lambda\phi^2}{\mu^2}
 \right).
\end{equation}
For homogeneous $\phi$, expanding to first order
in $\lambda$ yields
\begin{equation}
 \Gamma_1[\phi]
 =\Gamma_1[0]
 +\frac{\lambda\phi^2}{2}
 \operatorname{Tr}_{\mathcal N}
 (\Box_g+m^2)^{-1}
 +O(\lambda^2\phi^4).
\end{equation}
The reference scale $\mu$ is held fixed when
taking the derivative with respect to $m^2$.

In leading Weyl order, the correction to the
quadratic term in the effective potential is
\begin{equation}
 \delta M_H^2
 =\frac{2\lambda\hbar^4}{\mathcal V}
 \left.
 \frac{\partial\Gamma_1[0]}{\partial m^2}
 \right|_{\mu,\mathcal N,\widetilde L}
 =\frac{\lambda}{16\pi^2}
 \left[
 \widetilde L^2
 -m^2\ln\left(1+\frac{\widetilde L^2}{m^2}\right)
 \right].
\end{equation}

The entropy-selected equilibrium of Sec.~\ref{Sect:EntropyBound} fixes
$\widetilde L_{\rm eq}^2=\varepsilon_G^2 \stl_\star$.
Consequently,
\begin{equation}
 \boxed{
 \delta M_H^2(\mathcal N,\sm)
 =\frac{\lambda\varepsilon_G^2}{16\pi^2}
 \left[
 \stl_\star-\sm\ln\left(1+\frac{\stl_\star}{\sm}\right)
 \right].
 }
\end{equation}
This correction is positive and increases with
$\stl_\star$ at fixed $\sm$. However, the equilibrium
condition restricts $\stl_\star$ to the admissible interval
$0<\stl_\star<{\stl}_\infty(\sm)$.

A universal bound follows from the inequality
\begin{equation}
 \left[
 1-\frac{\ln(1+x)}{x}
 \right]^2<\tau(x),\qquad x>0,
\end{equation}
together with the equilibrium condition
${\stl_\star}^2\tau(\stl_\star/\sm)<4$. We obtain
\begin{equation}
 \boxed{
 0<\delta M_H^2<
 \frac{\lambda}{8\pi^2}\varepsilon_G^2.
 }
\end{equation}
The same bound holds for $\sm=0$.
Since $\varepsilon_G\simeq9.4\times10^{-3}\,
{\rm eV}$, the correction is negligible compared
with the observed Higgs mass squared.

The limiting cases are particularly instructive.
For a massless loop field, the equilibrium
solution gives
\begin{equation}
 \delta M_H^2(\mathcal N,0)
 =\frac{\lambda\varepsilon_G^2}{8\pi^2}
 \frac{\sr}
 {1+\sqrt{1+\sr^2}},
 \qquad
 \sr=\sqrt{\frac{\mathcal N}{\mathcal N_G}}.
\end{equation}
For $\sm\gg1$ at fixed $\sr$,
using
$\stl_\star=\sr-\sr^4/(6\sm)
+O(\sm^{-2})$ gives
\begin{equation}
 \delta M_H^2
 =\frac{\lambda\varepsilon_G^4}{32\pi^2m^2}
 \frac{\mathcal N}{\mathcal N_G}
 \left[1+O\left(\frac{\varepsilon_G^2}{m^2}\right)
 \right].
\end{equation}
Thus heavy fields decouple as $m^{-2}$ at
fixed cardinality ratio.

This suppression is not restricted to fixed
cardinality. Since
${\stl}_\infty
\sim(6\sm)^{1/3}$ for large $\sm$,
the admissible interval implies
\begin{equation}
 0<\delta M_H^2
 \lesssim
 \frac{\lambda\,6^{2/3}}{32\pi^2}
 \frac{\varepsilon_G^{8/3}}{m^{2/3}},
 \qquad m\gg\varepsilon_G.
\end{equation}
Hence the correction vanishes uniformly over
the admissible equilibrium solutions.

Similar results have been obtained by Ferrero and
Percacci~\cite{Ferrero:2024yvw}, who showed that
the one-loop corrections to the scalar mass and
quartic coupling become finite when the gravitational
background is determined self-consistently.
Their analysis uses a dimensionless spectral cutoff
and considers both symmetric and broken phases
of an interacting scalar field. Our approach is
different in that we work at fixed phase-space
cardinality and impose, in addition, the entropy
condition discussed in Sec.~\ref{Sect:EntropyBound}. This allows us
not only to establish finiteness, but also to
obtain an explicit upper bound on the mass
correction and to show that heavy fields decouple.
It should be stressed, however, that these
stronger results depend on our entropy assumption,
while the finiteness found in~\cite{Ferrero:2024yvw}
follows from the gravitational field equations
alone. Both results suggest that taking the
gravitational backreaction into account may
change considerably the usual understanding
of the hierarchy problem.

Our result here concerns the quadratic coefficient
of the homogeneous effective potential.
Establishing the stability of the physical
Higgs pole mass requires extending the analysis
to momentum-dependent two-point functions
and the complete Standard Model.

\section{Curvature corrections to the one-loop action}
\label{Sect:CurvatureCorrections}

So far we have evaluated the one-loop action at leading Weyl order.
There are curvature corrections to this result, which can be found by
expanding the restricted heat kernel in local momenta. For a minimally
coupled scalar field on a smooth four-dimensional Euclidean manifold,
the first correction is linear in the scalar curvature and takes the form

\begin{equation}
 \Gamma_1[g]=\frac12\operatorname{Tr}_{\mathcal N}
 \ln\left(
 \frac{\Box_g+m^2}{\mu^2}
 \right)=\mathcal N\,\gamma_1
 -\frac{\tL^2}{384\pi^2\hbar^2}
 \int_M d^4x\sqrt g\,R(g)\,
 \mathcal F\!\left(\frac{\tL^2}{m^2}\right)
 +\cdots,
 \label{eq:IX_curved_general}
\end{equation}
where $\gamma_1$ is the leading contribution given in Sec.~II and
\begin{equation}
 \mathcal F(z)=1+\frac{1}{1+z}
 -\frac{2}{z}\ln(1+z),
 \qquad \mathcal F(\infty)=1.
 \label{eq:IX_curved_F}
\end{equation}
The omitted terms include higher orders in the curvature. The formula
uses the same momentum-ball restriction as our leading calculation;
the subleading coefficients may depend on how this restriction is
implemented.

Let us specialize to the four-sphere of radius $L$. Since
$R=12/L^2$ and $\int_{S^4_L}d^4x\sqrt g\,R=32\pi^2 L^2$,
the fixed-cardinality relation $L^4\tL^4=12\mathcal N\hbar^4$
gives (see also discussion in the Appendix A, Eqs.~\eqref{app:A2-four-sphere}, \eqref{app:Gamma-curvature-sphere})
\begin{equation}
 \boxed{
 \Gamma_1=\mathcal N\,\gamma_1
 -\sqrt{\frac{\mathcal N}{12}}\,
 \mathcal F\!\left(\frac{\stl}{\sm}\right)
 +O(\mathcal N^0).
 }
 \label{eq:IX_curved_sphere}
\end{equation}
Thus, the first curvature correction grows as $\sqrt{\mathcal N}$,
while the leading action is proportional to $\mathcal N$.

In the notation of Sec.~V, the corrected stationary equation is
\begin{equation}
 \cE(\stl;\sr,\sm)
 =\frac{\stl^3}{2\sqrt{12\mathcal N}}\,
 \frac{\partial}{\partial\stl}
 \mathcal F\!\left(\frac{\stl}{\sm}\right)
 +O(\mathcal N^{-1}).
 \label{eq:IX_curved_stationarity}
\end{equation}
The correction is of relative order $\mathcal N^{-1/2}$.
Since the stationary root of Sec.~V is nondegenerate, its shift
is of the same order on the regular large-cardinality branch.
In the massless case $\mathcal F=1$ and the
$\sqrt{\mathcal N}$ term is independent of $\tL$ at fixed
$\mathcal N$; it therefore does not change the stationary equation
at this order. The corresponding corrections to the entropy-selection
condition are also subleading in the large-cardinality expansion, so
the leading behavior of $\Lambda_{\rm eff}$ and the scalar mass
correction obtained above is not changed. The precise subleading
entropy relation, and a fuller comparison of curvature-expansion and
spectral-cutoff prescriptions, will be discussed in a forthcoming paper.

\section{Conclusions}

In this paper we have explored the cosmological constant and hierarchy
problems from the viewpoint of the accessible phase-space cardinality.
Our starting point is the dimensionless volume $\mathcal N$ of the
accessible phase-space region, measured in units of elementary quantum
cells. On the Euclidean de Sitter four-sphere, fixing $\mathcal N$ relates
the spacetime radius $L$ to the momentum scale $\widetilde L$ through
$L^4\widetilde L^4=12\mathcal N\hbar^4$. Consequently, a variation of the
momentum scale at fixed cardinality also varies the geometry, and the
classical gravitational action participates in the scale variation.
This is the Cardinal RG=GR relation in the one-parameter gravitational
setting considered here.

At leading Weyl order the scalar one-loop effective action takes the
form $\Gamma_1=\mathcal N\gamma_1$. Including it in the gravitational
stationarity condition gives a unique regular, quantum-corrected de
Sitter solution for each positive cardinality and scalar mass. The
effective cosmological constant satisfies
$0<\Lambda_{\rm eff}<\Lambda$, and for
$\mathcal N=\mathcal N_G$ the relative decrease is at most
$3-2\sqrt{2}\simeq 0.172$ in the one-scalar model. At fixed
$G$, $\Lambda$ and scalar mass, the large-cardinality solution obeys
\begin{equation}
 \Lambda_{\rm eff}\simeq
 \Lambda\,\stl_{\infty}
 \sqrt{\frac{\mathcal N_G}{\mathcal N}},
 \qquad \sm=\frac{m^2}{\varepsilon_G^2},
 \qquad \mathcal N\gg\mathcal N_G.
\end{equation}
In particular, the curvature tends to zero as the universe grows.
The massless solution agrees with the self-consistent result of Becker
and Reuter, and the massive equations agree at leading Weyl order
\cite{Becker:2020mjl,Becker:2021pwo}. A different correlated limit,
with $G$ and $\mathcal N/\mathcal N_G$ fixed while both cardinalities
grow, instead gives $\Lambda_{\rm eff}\propto\mathcal N^{-1}$.
These are complementary statements about different families of
stationary solutions.

We have also studied the entropy of the coupled gravity--matter
system, using the thermodynamic prescription $S=-\Gamma_\star$ and a
holographic bound evaluated on the stationary geometry. Entropy
maximization selects the cardinality-dependent normalization of the
one-loop determinant and gives
$\widetilde L_{\rm eq}^{\,2}=\varepsilon^2$. At this equilibrium the
bound is saturated, and the effective cosmological constant and the
maximum entropy satisfy the exact relation
\begin{equation}
\Lambda_{\rm eff} S_{\max}=\frac{3\pi}{\hbar G}.
\end{equation}
For fixed gravitational parameters, $S_{\max}\propto\sqrt{\mathcal N}$
at large cardinality. This extends the connection between gravitational
entropy and vacuum energy discussed in our earlier work
\cite{Freidel:2022ryr,Freidel:2023ytq}, while the entropy principle
provides the additional selection of the one-loop stationary solution.

The same construction gives a bounded correction to the mass parameter
of a homogeneous Higgs-like scalar coupled to a quantum scalar field.
For a positive coupling $\lambda$, the one-loop result satisfies
\begin{equation}
0<\delta M_H^2<\frac{\lambda\varepsilon_G^2}{8\pi^2}.
\end{equation}
It is controlled by the small gravitational scale $\varepsilon_G$ and
tends to zero for a heavy field throughout the admissible range of
entropy-selected equilibria. The result concerns the mass parameter
in the homogeneous effective potential; a physical Higgs pole mass
requires the momentum-dependent two-point function.

The relation with conventional quantum field theory becomes explicit
in the double-scaling limit $G,\Lambda\to0$ at fixed $\Lambda/G$ and
$\mathcal N/\mathcal N_G$. The spacetime curvature then vanishes while
a finite momentum scale survives, and the matter contribution per unit
volume reduces to the familiar flat-space one-loop cutoff integral.
We have also considered the first curvature corrections to the
restricted determinant. On the four-sphere they enter $\Gamma_1$ at
order $\sqrt{\mathcal N}$, compared with its leading
$O(\mathcal N)$ term, and give subleading changes to the regular
large-cardinality stationary solution.

It is tempting to compare the cosmological constant problem with the
ultraviolet catastrophe of black-body radiation. In the latter, a new
understanding of how states contribute to the energy changed the
physical picture. Here the relevant change is to account for the
gravity-dependent accessible phase space already in the counting of
quantum fluctuations. The analogy is heuristic, but it illustrates
our central point: the ultraviolet momentum range and the infrared
geometry belong to the same gravitational system.

One of the most important results is the explicit demonstration of the radiative stability of both the cosmological constant and the scalar (Higgs) mass. Another important result is the existence of a
double-scaling limit in which gravity decouples
and the background becomes flat, while the
momentum cutoff remains finite. In this limit
we recover the standard one-loop expression
of cutoff QFT, with the momentum scale related to
gravitational parameters and the cardinality,
rather than being introduced independently.

However, unlike the cosmological constant, which we have directly tied to geometric mean of the cosmological horizon scale and the Planck scale, the mass of the scalar field was not directly evaluated in this paper.
Nevertheless, one could also present an argument which leads to the value of the scalar fields as a geometric mean between the vacuum energy and the Planck scale~\cite{hubsch2024quantum, Hubsch:2026lot}. This is compatible with spectral counting
once the next to leading order terms are taking into account and it will be discussed elsewhere.

There are several further directions which seem natural. The approach should be extended to fermions, vector fields and non-Abelian gauge theories, and eventually to the full Standard Model. It would also be important to understand the construction beyond one loop.

The discussion should also be extended beyond the highly symmetric de Sitter background, and eventually to Lorentzian spacetime, including black hole backgrounds. In this respect it would be interesting to understand more directly the relation of the present RG=GR picture to other places where geometry and renormalization are tied together, for example the sigma-model argument underlying string theory and the holographic renormalization group \cite{Friedan:1980jm,Susskind:1998dq,deBoer:1999tgo,Strominger:2001gp, Balasubramanian:2001nb}.

The main lesson of the present paper is rather simple. Once the accessible phase space is characterized by a finite dimensionless cardinality, spacetime size and momentum range are no longer independent. This changes the way matter loops couple to the gravitational background and, in the examples considered here, removes the large radiative instability of the cosmological constant and of a scalar mass. Whether this idea can be developed into a general framework for quantum field theory in a dynamical spacetime remains an open question.


\section*{Acknowledgment}
We thank  P.~Berglund,  J.~Erlich, A.~Geraci, S.~Hellerman, T.~H\"{u}bsch, S.~Iso, D.~Mattingly, R.~ Percacci, M.~Petropoulos, A.~Sagnotti, H.~Steinacker, T.~Takeuchi, E.~Verlinde and K.~Zurek for
discussions. JKG and DM thank  Perimeter  Institute  for  hospitality. LF, JKG and DM thank the Institute for Fundamental Physics of SISSA, Italy, and its director Stefano Liberati for hospitality during a week in May of 2025. JKG is supported by National Science Centre (Poland) grant 2025/59/B/ST2/00095.
DM is supported in part by the U.S. Department of Energy under contract DE-SC0020262.
Research at Perimeter Institute for Theoretical Physics is supported in part by the Government of Canada through NSERC and by the Province of Ontario through MRI. This work contributes to the European Union COST Action CA23130 {\it ``Bridging high and low energies in search of quantum gravity (BridgeQG)''}.

\textbf{Use of generative AI}. ChatGPT (OpenAI) was used during the preparation of this manuscript to assist with language editing and organization of the text, and as an aid in checking some intermediate calculations. All results and conclusions were independently verified by the authors, who take full responsibility for the final manuscript.


\appendix

\section{Heat Kernel and spectral counting}\label{app:A}

The aim of this appendix is to clarify how spectral methods relate to the
cardinal construction used in the main text. In Sec.~II we define the
cardinality $\cN$ from the volume of the accessible phase-space region
and calculate the one-loop action using the corresponding restricted trace.
Here we first compare this geometrical counting with the spectral count of
a Laplace-type operator.

We work in Euclidean signature on a compact smooth four-dimensional Riemannian manifold $M$ without boundary. We consider the differential operator and curvature
\begin{equation}\label{eq:conventions}
P=-\nabla^2+U, \qquad
F_{\mu\nu}=[\nabla_\mu,\nabla_\nu],
\end{equation}
where $\nabla_\mu$ includes the connection on the bundle on which $P$ acts.
The operator $P$ has dimensions of inverse length squared. Its associated
operator with dimensions of momentum squared is $\hbar^2P$.
Thus, if $\tL$ is a momentum scale, the dimensionless operator entering
the spectral function is $\hbar^2P/\tL^2$. This restores $\hbar$
in the formulas below, while the heat-kernel coefficients are those of $P$.
For a scalar with physical mass $m$, for example, the potential contains
$U=m^2/\hbar^2$ (possibly with further curvature terms).

The associated heat trace has the small-$s$ expansion
\begin{equation}\label{eq:trace-hk}
\Tr e^{-sP} \sim\frac1{(4\pi s)^2}
\left[A_0+sA_2+s^2A_4+s^3 A_6+\cdots\right],
\qquad
A_{2n}:=\int_M\tr a_{2n}(x) \rd \nu_x.
\end{equation}
The RHS is an asymptotic expansion near $s=0$, and the integrated coefficient $A_{2n}$ has
mass dimension $2n-4$.
The coefficients $a_{2n}$ are the standard Seeley--DeWitt coefficients, see \cite{Vassilevich:2003xt, Avramidi:1997jy}.
The first ones are
\begin{align}\label{eq:a0a2}
a_0 &=\one,\qquad a_2=\frac16R\one-U, \\
a_4 &=\frac12\left(\frac16R\one-U\right)^2
+\frac1{12}F_{\mu\nu}F^{\mu\nu}
+\frac1{180}\left(R_{\mu\nu\rho\sigma}R^{\mu\nu\rho\sigma} - R_{\mu\nu}R^{\mu\nu}\right)\one
\cr
&
+\frac16\nabla^2\left(\frac15 R \one-U\right).
\end{align}
On a closed manifold, the integral of the last term vanishes.

The spectral regularization of the trace is given by the choice of a smooth function $F$ on $\mathbb{R}^+$ which is chosen to decay exponentially when $x\to\infty$ and is such that $F(0)=1$.
Let $\lambda_j$ be the eigenvalues of $P$, which we assume to be
non-negative, counted with their multiplicities. The sharp spectral cutoff
function is given by $F_{\rm sharp}(x)=\Theta(1-x)$, $x=\hbar^2\lambda_j/\tL^2$, selects the eigenvalues satisfying $\hbar^2\lambda_j\leq\tL^2$. Then
\begin{equation}
N_{\rm sharp}(\tL)
=\sum_j\Theta\!\left(1-\frac{\hbar^2\lambda_j}{\tL^2}\right)
\end{equation}
counts eigenvalues below the ultraviolet scale $\tL$.

Notice that
$N_{\rm sharp}(\tL)$ is a staircase, since it jumps when
$\tL$ crosses an eigenvalue. Its leading Weyl term is
\begin{equation}
N_{\rm Weyl}(\tL)
=\frac{A_0}{32\pi^2}\frac{\tL^4}{\hbar^4},
\end{equation}
and the difference between the exact counting function and this leading
term is the remainder
\begin{equation}
R_{\rm sharp}(\tL)
:=N_{\rm sharp}(\tL)-N_{\rm Weyl}(\tL).
\end{equation}
The remainder depends on the positions and degeneracies of the eigenvalues.
Thus the heat-kernel coefficients do not in general unambiguously determine
the subleading terms of a pointwise expansion of the sharp counting function.

This is why we first consider a smooth spectral function $F$. Its spectral
count has an asymptotic expansion whose local coefficients can be computed
from the heat kernel.
Following \cite{oleszczuk1994symmetry, Chamseddine:2009drp, Bilal:2013iva} we therefore assume that $F$ can be written as the Laplace transform of $\hat{F}$
\be
F(x) =\int_0^\infty \hat{F}(s) e^{-sx}\rd s,
\ee
and $n=1,2,\ldots$ we denote
\begin{align}
F_n &:= \int_0^\infty \frac{\hat{F}(s)}{s^n} \rd s
= -\frac1{n!} \int_0^\infty x^n F'(x) \rd x\\
F_{-n} &:=
\int_0^\infty \hat{F}(s) {s^n} \rd s
= (-1)^n F^{(n)}(0).
\end{align}
We also write $F_0:=F(0)=\int_0^\infty\hat F(s)\rd s$.  In particular, $F_1=\int_0^\infty F(x)\rd x$ and $F_2=\int_0^\infty xF(x)\rd x$. The equalities above assume sufficient decay and regularity for the integrals and integrations by parts to exist.

This shows that
\begin{align}
N_F(\tL) &:= \Tr\left(
F\left(
{\frac{\hbar^2P}{\tL^2}}
\right)
\right) =
\int_0^\infty \rd s\,\hat{F}(s)\,\Tr\left(e^{-\frac{s\hbar^2P}{\tL^2}}\right)\\
&=\frac{1}{(4\pi)^2}
\left(
\frac{A_0\tL^4}{\hbar^4}F_2
+\frac{A_2\tL^2}{\hbar^2}F_1
+A_4F_0+\frac{\hbar^2 A_6}{\tL^2}F_{-1}
+\cdots
\right).
\label{bf-heat-conventions}
\end{align}
We can separate this evaluation into the local dominant contribution and the rest which vanishes when ${\tL}\to\infty$.
\be
N^{\rm loc}_P(\tL):=\frac{1}{(4\pi)^2}\left(
\frac{A_0\tL^4}{\hbar^4}F_2
+\frac{A_2\tL^2}{\hbar^2}F_1+A_4F_0\right).
\ee

To compare with the geometrical cardinality we need only normalize the
leading term by choosing $F_2=\frac12$. Since $F(0)=1$, we also have
$F_0=1$, whereas $F_1$ remains dependent on the choice of smoothing.
It is interesting to compare these coefficients with the ones corresponding
to the sharp cutoff. The function $F_{\rm sharp}(x)=\Theta(1-x)$ has the moments
\be
F_2^{\rm sharp}=\frac12,\qquad
F_1^{\rm sharp}=F_0^{\rm sharp}=1,
\ee
which means that
\be
\int_0^\infty x F_{\rm sharp}(x)\rd x=\frac12,
\qquad
\int_0^\infty F_{\rm sharp}(x)\rd x=1,
\qquad F_{\rm sharp}(0)=1.
\ee
Thus if we also impose $F_1=1$, we obtain three conditions which cannot
hold exactly for a smooth monotonically decreasing function $F$.
Note that this restriction of the coefficients is such that
\be
\int_0^\infty (x-1)^2F'(x)\rd x=0,
\ee
and therefore the only function $F$ which is monotonically decreasing on $\mathbb R^+$ and satisfies these conditions is the Heaviside step function $F(x)=\Theta(1-x)$ which is not smooth.
We can nevertheless take a limiting family of smooth cutoffs approaching
$\Theta(1-x)$, and in this way recover the sharp values of their moments.
We stress again that this is only a limit of the coefficients of the
smoothed expansion, and not an asymptotic expansion of the exact sharp
spectral counting function.

To express the expansion in terms of the cardinality, we use the volume
$\mathcal V=\int_M\rd\nu_x$ and the momentum ball of radius $\tL$.
In four dimensions
\be
\mathcal N
=\frac{\mathcal V}{(2\pi\hbar)^4}
\operatorname{Vol}(B^4_{\tL})
=\frac{\mathcal V\tL^4}{32\pi^2\hbar^4},
\qquad
\frac{\tL^2}{\hbar^2}
=\frac{4\pi\sqrt{2\mathcal N}}{\sqrt{\mathcal V}}.
\ee
For a bundle of rank $N_E$, let $\tau_E=\frac1{N_E}\tr_E$ and write
\begin{align}
W_E&:=U-\frac16R\one,\qquad W:=\tau_E(W_E),\\
C&:=\frac12\tau_E(W_E^2)+\frac1{12}\tau_E(F_{\mu\nu}F^{\mu\nu})
+\frac1{180}\left(R_{\mu\nu\rho\sigma}R^{\mu\nu\rho\sigma}-R_{\mu\nu}R^{\mu\nu}\right).
\end{align}
Then, since there is no boundary,
\be
A_0=N_E\mathcal V,\qquad
A_2=-N_E\int_M W\rd\nu_x,\qquad
A_4=N_E\int_M C\rd\nu_x.
\ee
Thus the expansion can be written in terms of $\cN$ as
\be
\frac{N_P^{\rm loc}(\mathcal N)}{N_E}
=2F_2\mathcal N
-\frac{F_1\sqrt{2\mathcal N}}{4\pi\sqrt{\mathcal V}}
\int_M W\rd\nu_x
+\frac{F_0}{(4\pi)^2}\int_M C\rd\nu_x.
\ee
In the sharp-profile limit $F_2\to\frac12$, $F_1\to1$, $F_0\to1$,
this becomes
\be
\frac{N_P^{\rm loc}(\mathcal N)}{N_E}
=\mathcal N
-\frac{\sqrt{2\mathcal N}}{4\pi\sqrt{\mathcal V}}
\int_M W\rd\nu_x
+\frac{1}{(4\pi)^2}\int_M C\rd\nu_x.
\ee
This is the sharp-profile limit of the local coefficients used in
Section~II. For a fixed smooth $F$ with $F_2=1/2$, the leading term is
the geometrical cardinality, whereas the subleading coefficients retain
their dependence on the choice of smoothing.

\subsection{Spectral representation of the restricted one-loop effective action}\label{app:restricted-spectral-action}

The spectral counting function considered above is related to the cardinality
at leading Weyl order.  We now explain how the same heat-kernel techniques
can be applied to the one-loop effective action used in Sec.~II.  There is
an important distinction from a spectral cutoff on the full operator $P$:
the geometrical restriction in the main calculation is on the momentum
region and is independent of the mass $m$.  For a minimally coupled scalar
let
\be\label{app:kinetic-Q}
Q:=-\hbar^2\nabla^2=\Box_g,\qquad D_m:=Q+m^2.
\ee
Thus $Q$ has dimensions of momentum squared, while the determinant
contains $D_m$.

Recall that the ordinary one-loop effective action of a free
scalar field is given by the logarithmic determinant of its
quadratic operator $D_m=Q+m^2$. In the eigenbasis of $Q$,
with $Q\phi_j=q_j\phi_j$, this determinant is a sum of
$\ln[(q_j+m^2)/\mu^2]$ over all eigenvalues.
Restricting the sum to eigenvalues $q_j\leq\tL^2$
amounts to inserting the spectral projector
$\Theta(1-Q/\tL^2)$ into the trace.
A smooth spectral restriction is obtained by replacing
this projector with a smooth, sufficiently rapidly decreasing spectral function $F(Q/\tL^2)$, of the type discussed above.
This motivates the following spectral version of the
restricted one-loop effective action:
\be\label{app:Gamma-F-eps}
\Gamma_{1,F}(\tL;m,\mu)
:=\frac12\Tr\!\left[
F\!\left(\frac{Q}{\tL^2}\right)
\ln\!\left(\frac{Q+m^2}{\mu^2}\right)
\right].
\ee
The two scales $\tL$ and $\mu$ play different
roles: $\tL$ determines the restriction of the trace and $\mu$
normalizes the logarithm.
In Eq.~\eqref{app:Gamma-F-eps} the operator inside $F$ is $Q$, not $D_m$.  The distinction matters
in particular for $m\gg\tL$: a cutoff on $D_m$ would remove the massive
field from the spectral count, whereas the phase-space region defining
$\cN$ does not shrink when $m$ is increased.    On a general curved background Eq.~\eqref{app:Gamma-F-eps}
does not need not coincide with the geometrically restricted trace $\Tr_{\cN}$
beyond leading Weyl order.

For $m>0$, the logarithm admits the Schwinger representation, and hence
\begin{align}\label{app:Gamma-F-propertime}
\Gamma_{1,F}
=\frac12\int_0^\infty\frac{\rd\tau}{\tau}\Bigg\{
&e^{-\mu^2\tau}\Tr F\!\left(\frac{Q}{\tL^2}\right)
-e^{-m^2\tau}\Tr\!\left[
F\!\left(\frac{Q}{\tL^2}\right)e^{-\tau Q}
\right]\Bigg\}.
\end{align}
For a smooth $F$ and positive $m$ the spectral traces are well defined
on compact $M$ and the cancellation in the braces makes the proper-time
integral finite at $\tau=0$.  Using the Laplace representation of the function $F$ introduced above,
we have the exact identity
\be\label{app:Gamma-F-heatbridge}
\Tr\!\left[
F\!\left(\frac{Q}{\tL^2}\right)e^{-\tau Q}
\right]
=\int_0^\infty \rd s\,\hat F(s)\,
\Tr\exp\!\left[-\left(\tau+\frac{s}{\tL^2}\right)Q\right].
\ee

Equations (A25) and (A26) connect the spectrally restricted
determinant to the heat kernel. We can now show explicitly
that they reproduce the one-loop effective action derived
in Sec.~II. At leading Weyl order in four dimensions,
\begin{align}
\Tr e^{-tQ}\simeq
\frac{\mathcal V}{(4\pi\hbar^2 t)^2}.
\end{align}
Substituting this expression into Eq.~(A26) and using
\begin{align}
\frac{1}{(\tau+s/\tL^2)^2}
=\int_0^\infty du\,u\,
e^{-u(\tau+s/\tL^2)},
\end{align}
we obtain
\begin{align}
\Tr\left[F(Q/\tL^2)e^{-\tau Q}\right]
&\simeq
\frac{\mathcal V}{16\pi^2\hbar^4}
\int_0^\infty du\,u\,
F(u/\tL^2)e^{-\tau u}
=2\cN\int_0^\infty dx\,xF(x)
e^{-\tau\tL^2x},
\end{align}
where we used
$\cN=\mathcal{V}\tL^4/(32\pi^2\hbar^4)$.

For the sharp profile $F(x)=\Theta(1-x)$, this gives
\begin{align}
\Tr\left[F(Q/\tL^2)e^{-\tau Q}\right]
\simeq
\frac{2\cN}{z^2}
\left[1-(1+z)e^{-z}\right],
\qquad z=\tau\tL^2,
\end{align}
while $\Tr F(Q/\tL^2)\simeq\cN$.
Inserting these expressions into Eq.~(A25), and using
the prescription
$\mu^2=m^2+\varepsilon^2$, we find
\begin{align}
\Gamma_{1,F}^{\rm Weyl}
=\frac{\cN}{2}\int_0^\infty
\frac{d\tau}{\tau}\,e^{-m^2\tau}
\left[
e^{-\varepsilon^2\tau}
-\frac{2}{z^2}
\left(1-(1+z)e^{-z}\right)
\right].
\end{align}
The expression in square brackets, together with the
overall factor $\cN e^{-m^2\tau}$, is precisely the
proper-time function $Z_{\cN}(\tau)$ of Eq.~(35).

Performing the proper-time integration, or equivalently
using the Frullani identity before the momentum
integration, gives
\begin{align}
\begin{aligned}
\Gamma_{1,F}^{\rm Weyl}
&=\cN\int_0^1 dx\,x\,
\ln\left(\frac{\tL^2x+m^2}
{m^2+\varepsilon^2}\right)=\cN\gamma_1,
\end{aligned}
\end{align}
where $\gamma_1$ is exactly the expression given in
Eq.~(37). We have thus recovered Eqs.~(36)--(37)
directly from the heat-kernel representation of the
spectrally restricted determinant.

This equivalence holds for the sharp profile at leading
Weyl order. It does not imply an exact identification
of the spectral restriction and the geometrical
restriction on a curved manifold.

The massless case deserves a qualification.  On a compact four-sphere
$Q$ has a constant zero mode; its contribution to the exact logarithm
in~\eqref{app:Gamma-F-eps} is singular at $m=0$.  One must then remove
the zero mode from the spectral determinant or use an infrared
prescription.  This does not affect the leading phase-space integral,
where the single point $p=0$ has zero measure, but it can affect
subleading spectral terms.

\subsection{Leading Weyl limit and curvature terms}

To compare~\eqref{app:Gamma-F-eps} with the main calculation we use
the principal symbol of $Q$, which is $p^2$.  At leading Weyl order,
\begin{align}\label{app:Gamma-F-Weyl}
\Gamma_{1,F}
&\simeq\frac{\mathcal V}{2}\int_{\mathbb R^4}
\frac{\rd^4p}{(2\pi\hbar)^4}
F\!\left(\frac{p^2}{\tL^2}\right)
\ln\!\left(\frac{p^2+m^2}{\mu^2}\right)
\nonumber\\
&=\mathcal N\int_0^\infty \rd x\,xF(x)
\ln\!\left(\frac{\tL^2x+m^2}{\mu^2}\right),
\qquad
\mathcal N=\frac{\mathcal V\tL^4}{32\pi^2\hbar^4}.
\end{align}
For $F_{\rm sharp}(x)=\Theta(1-x)$ this reduces to
\be\label{app:Gamma-F-sharp-leading}
\Gamma_{1,F}^{\rm Weyl}
=\frac{\mathcal V}{2}
\int_{|p|\leq\tL}\frac{\rd^4p}{(2\pi\hbar)^4}
\ln\!\left(\frac{p^2+m^2}{\mu^2}\right)
=\mathcal N\,\gamma_1(\tL;m,\mu).
\ee
This is precisely the integral used in Sec.~II, where the choice
$\mu^2=m^2+\varepsilon^2$ is made after defining the trace.

The possible subleading contributions can also be displayed.  Write
\be\label{app:weighted-W}
\mathscr W(u):=
F\!\left(\frac{u}{\tL^2}\right)
\ln\!\left(\frac{u+m^2}{\mu^2}\right).
\ee
For smooth $F$, $m>0$, and in a regime in which a local heat-kernel
expansion of the corresponding spectral trace is justified, its first
terms take the form
\begin{align}\label{app:Gamma-F-local}
\Gamma_{1,F}^{\rm loc}\sim\frac{1}{2(4\pi)^2}\Bigg[
&\frac{A_0^{(0)}}{\hbar^4}
\int_0^\infty\rd u\,u\,\mathscr W(u)
+\frac{A_2^{(0)}}{\hbar^2}
\int_0^\infty\rd u\,\mathscr W(u)
+\cdots\Bigg],
\end{align}
where $A_{2n}^{(0)}$ are the Seeley--DeWitt coefficients of
$P_0:=Q/\hbar^2=-\nabla^2$, i.e.\ with the scalar mass excluded
from the operator defining the spectral restriction.  The first term
is~\eqref{app:Gamma-F-Weyl}.  The second is a local curvature
contribution, distinct from the $O(\sqrt{\cN})$ correction to the
\emph{spectral count} derived above.  Terms beyond those shown,
including possible nonlocal spectral contributions, require a separate
analysis.  In particular, taking a sharp-profile limit does not
establish the validity of the subleading local expansion for the
exact staircase spectrum.

For a minimally coupled scalar on the round $S_L^4$, the second
coefficient is
\be\label{app:A2-four-sphere}
A_2^{(0)}=\frac16\int_{S_L^4}R\,\rd\nu_x
=\frac{2\mathcal V}{L^2},
\qquad R=\frac{12}{L^2}.
\ee
Thus the term proportional to $A_2^{(0)}$ is formally
\begin{align}\label{app:Gamma-curvature-sphere}
\Gamma_{1,F}^{(R)}
&=\frac{\mathcal V}{16\pi^2 L^2\hbar^2}
\int_0^\infty\rd u\,
F\!\left(\frac{u}{\tL^2}\right)
\ln\!\left(\frac{u+m^2}{\mu^2}\right)
=\sqrt{\frac{\mathcal N}{3}}
\int_0^\infty\rd x\,F(x)
\ln\!\left(\frac{\tL^2x+m^2}{\mu^2}\right),
\end{align}
where in the last line we used $L^4\tL^4=12\cN\hbar^4$.
The displayed terms suggest an order-$\cN$ leading contribution and
a possible order-$\sqrt{\cN}$ correction when the dimensionless
mass ratios and profile are held fixed.  They should not be inserted
into the stationary equation of the main text without first
specifying the spectral restriction and controlling its curvature
and regulator dependence.

One special feature of the four-sphere is worth noting.  The exact
eigenvalues of $Q$ are
\be\label{app:Q-spectrum-sphere}
q_\ell=\frac{\hbar^2}{L^2}\,\ell(\ell+3),
\qquad
\frac{q_\ell}{\tL^2}
=\frac{\ell(\ell+3)}{\sqrt{12\cN}},
\qquad \ell=0,1,\ldots.
\ee
Hence the spectral \emph{restriction} $F(Q/\tL^2)$ is constant
along the family of four-spheres at fixed $\cN$ (after the natural
identification of their spherical harmonics).  Its exact spectral
count is nevertheless not equal to the geometrical $\cN$ in
general.

Finally, at fixed metric, $\tL$, and $\mu$, differentiation of
\eqref{app:Gamma-F-eps} with respect to the mass gives
\be\label{app:mass-deriv}
\left.\frac{\pa\Gamma_{1,F}}{\pa m^2}\right|_{g,\tL,\mu}
=\frac12\Tr\!\left[
F\!\left(\frac{Q}{\tL^2}\right)
\frac{1}{Q+m^2}\right].
\ee
The normalization prescription is to be imposed only after this
derivative, as in the calculation of the homogeneous effective
potential in the main text.

\subsection{
An alternative spectral flow}

Given a positive Laplace-type operator $P$ and the same
profile $F$, define
\be\label{app:Gamma-H-definition}
H(x):=-\int_x^\infty\frac{F(y)}{y}\,\rd y,
\qquad H'(x)=\frac{F(x)}{x},
\qquad
\Gamma_H(\tL):=\frac12\Tr H\!\left(
\frac{\hbar^2P}{\tL^2}\right).
\ee
We assume that $P$ has no zero modes, or that they have been
removed.  At low positive eigenvalues, $H$ behaves as
\be\label{app:H-low-eigenvalues}
H(x)=\ln x+c_F+o(1),\qquad
c_F=-\int_0^1\frac{F(y)-1}{y}\,\rd y
-\int_1^\infty\frac{F(y)}{y}\,\rd y.
\ee
The corresponding normalization scale is
$\mu_H^2=\tL^2e^{-c_F}$, and therefore varies
with the spectral cutoff.  Moreover, at fixed $\tL$,
\be\label{app:H-variation}
\delta\Gamma_H
=\frac12\Tr\!\left[
F\!\left(\frac{\hbar^2P}{\tL^2}\right)
P^{-1}\delta P\right].
\ee
The scale dependence at fixed $P$ satisfies the exact identity
\be\label{app:H-flow-identity}
\tL\frac{\pa\Gamma_H}{\pa\tL}
=-\Tr F\!\left(\frac{\hbar^2P}{\tL^2}\right)
=-N_F(\tL).
\ee
Inserting~\eqref{bf-heat-conventions} and integrating the scale
derivative yields the local cutoff-dependent terms
\begin{align}\label{bf-four-dimensional-expansion}
\Gamma_H(\tL)\sim-\frac{1}{(4\pi)^2}\Bigg[
&\frac{A_0F_2}{4}\frac{\tL^4}{\hbar^4}
+\frac{A_2F_1}{2}\frac{\tL^2}{\hbar^2}
+A_4F_0\ln\!\left(\frac{\tL}{\varepsilon_0}\right)
\Bigg]+C_H(P,\mu_0)
+O\!\left(\frac{\hbar^2}{\tL^2}\right).
\end{align}
Here $\mu_0$ is an arbitrary reference momentum scale and
$C_H$ is independent of $\tL$ and need not be local.
In the sharp-profile limit
\be\label{app:H-sharp}
H(x)=\Theta(1-x)\ln x,
\qquad
\Gamma_H(\tL)
=\frac12\Tr\!\left[
\Theta(\tL^2-\hbar^2P)
\ln\!\left(\frac{\hbar^2P}{\tL^2}\right)
\right].
\ee
Thus $\tL$ fixes both the spectral threshold and the
normalization of the logarithm.  This is \emph{not} the action
\eqref{app:Gamma-F-eps}: in that action the normalization
$\mu$ is independent of the kinetic restriction, and
$P$ may include a potential or mass term.  The exact identity
\eqref{app:H-flow-identity} is therefore a property of a different
spectral construction, not the Cardinal RG=GR equation of the main text.

\section{ Calculation of the one-loop effective action: Technical details} \label{AppA}


Here we present a few technical details of the computation of the one-loop effective action
\begin{align}
    \Gamma_1{}_{(N,\mu)}(m) :=1/2 \ln\det[\mu^{-2}(\Box_g + m^2)]
= \frac12\int_0^{\infty} \frac{\rd \tau}{ \tau} Z_{(N,\mu)}(m;\tau).
\end{align}
We focus on the phenomenologically relevant case $d=4$. In this case, the cardinal number is given by
\be
 \mathcal{N}= \frac{(L\tL)^4}{12 \hbar^4},
\ee
and the
 partition function  becomes
\begin{align}
Z_{(N,\mu)}(m;\tau) &
 =\mathcal{N}\left[ e^{-\mu^2\tau}  - e^{-m^2\tau} \left[\int_{|p|<\tL}  e^{-p^2  \tau}\, \frac{\rd^4 p}{\tilde{\mathcal{V}}} \right]\right] = \mathcal{N}\left[ e^{-\mu^2\tau}- e^{- m^2\tau}
 2 E_2(\tau \tL^2)
 \right],
\end{align}
where the function entering the partition function is defined as
\begin{align} \label{incompletegamma}
E_n(z):= \frac{\gamma_n(z)}{z^n} :=z^{-n}\int_0^z u^{n-1} e^{-u} d u  = \frac{(n-1)!}{z^n} \left[1-e^{-z} \sum_{k=0}^{n-1} \frac{z^k}{k!}\right], \quad \text{for}\quad n\in \mathbb{N}^*,
\end{align}
and it represents the ratio of the lower incomplete gamma function of order $n$ with $z^n$. This means that
the one-loop effective action is
\begin{align}\label{modefaction}
    \Gamma_1{}_{(N,\mu)}  &=\frac{\mathcal{N} }2 \int_0^{\infty} \frac{d \tau}{ \tau} \left[ e^{-\mu^2\tau}- 2\frac{ e^{-m^2\tau}}{\tL^4 \tau^2} + 2 \frac{e^{-(\tL^2+m^2)\tau}}{\tL^4 \tau^2} \qty(1+ \tL^2\tau)  \right].
\end{align}

\paragraph{Exact evaluation}
The functions $E_n$ satisfy the recursion relation
\be
 \frac{n}{z} E_n(z) = \frac{e^{-z}}{z} + E_{n+1}(z), \qquad  E_1(z)=\frac{1-e^{-z}}{z}.
\ee
We can use this expansion to split the one-loop effective action. The 1-loop effective action  $\Gamma_1{}_{(N,\varepsilon)}(m) $  is a sum of  UV  and IR contributions
$\Gamma_1{}_{(N,\mu)} = \Gamma^{\text{UV}}_{(N,\mu)}+ \Gamma^{\text{IR}}_{(N,\mu)}$ where
\begin{align}
  \Gamma^{\text{UV}}_{(N,\mu)}&= \frac{\mathcal{N} }2 \int_0^{\infty} \frac{d \tau}{ \tau} \left[ e^{-\mu^2\tau}- e^{-(\tL^2+m^2)\tau}  \right]= \frac{\mathcal{N}}2 \ln\left(\frac{\tL^2+m^2}{\mu^2}\right)\\
  \Gamma^{\text{IR}}_{(N,\mu)} &= -\frac{\mathcal{N}\tL^2 }2 \int_0^{\infty} \rd \tau \, e^{- m^2\tau} E_3(\tau\tL^2).
\end{align}
The UV contribution contains  the $e^{-\mu^2\tau}$ term which is required to have the argument of $\ln\det$ dimensionless in the definition of one loop effective action. Remarkably the presence of this term is all that is needed to make the integral \eqref{modefaction} finite. The IR contribution is manifestly finite since $E_3(z)$ is regular around $z=0$ and decreases as $2/z^3$ for large $z$ (with an additional exponential factor $e^{-m^2\tau}$ for $m>0$). Its analytical expansion is
given by
\be
E_3(\tau \tL^2) =  2 e^{- (\tau \tL^2)}
\sum_{n=0}^\infty\frac{ (\tau \tL^2)^n}{(n+3)!}.
\ee
Therefore the integral is given by
\begin{align}
 \Gamma^{\text{IR}}_{(N,\mu)} &=-\mathcal{N}  \sum_{n=0}^{\infty} \frac{\tL^{2(n+1)}}{(n+3)!} \int_0^{\infty} \rd \tau \, \tau^{n} e^{- (\tL^2+m^2)\tau} \nonumber\\
 &= -\mathcal{N} \sum_{n=0}^{\infty} \frac{n!}{(n+3)!}\left(\frac{\tL^{2}}{\tL^2+m^2}\right)^{n+1} = -\frac{\mathcal{N} }2
 \left(\frac12 - \frac{m^2}{\tL^2} + \frac{m^4}{\tL^4} \ln\left(\frac{\tL^2+m^2 }{m^2}\right)   \right).
\end{align}
In the last equality we used the following reasoning:
Let us define
$$
F(x) =\sum_{n=0}^\infty \frac{x^{n+1}}{(n+3)(n+2)(n+1)}
$$
and $G(x) = x^2 F(x)$, where $G$  is such  that $ G'''(x) =\frac{1}{(1-x)}$ and $G(0)=G'(0)=G''(0)=0$. Therefore
\be
G(x) = -\frac12
\left((1-x)^2 \ln(1-x) -\frac32 x^2 + x \right).
\ee
In total we find the expression cited in the main text \eqref{Gammaex}
\begin{align}
\label{BGammaex}
\boxed{
\Gamma_1 =
\frac{\mathcal{N}}2 \left[\ln\left(\frac{\tL^2+m^2}{\mu^2}\right)
-
 \frac12 + \frac{m^2}{\tL^2} - \frac{m^4}{\tL^4} \ln\left(\frac{\tL^2+m^2  }{m^2}\right)
 \right]
 }.
\end{align}
We can expand this expression for large $\tL/m$ as
\begin{align}
\Gamma_1 &=\frac{\mathcal N}{2}\left[
\ln\!\left(\frac{\tL^{2}}{\mu^{2}}\right)
-\frac12
+\frac{2m^{2}}{\tL^{2}}
-\frac{m^{4}}{\tL^{4}}\left(\ln\!\frac{\tL^{2}}{m^{2}}+\frac12\right)
+O\!\left(\frac{m^{6}}{\tL^{6}}\right)
\right].
\end{align}

\subsection{Dimensional regularization:}
Although the complete proper-time integral in eq.\ \eqref{modefaction} is finite, its individual terms are singular at $\tau=0$ and therefore cannot be integrated separately without an auxiliary prescription. As an independent check of the evaluation above, we introduce an analytic regulator $\alpha$, together with an auxiliary scale $M_{\rm DR}$, and perform the individual integrals for values of $\alpha$ for which they are well defined. The regulator is removed only after all terms have been combined; consequently, both the poles at $\alpha=0$ and the dependence on the auxiliary scale $M_{\rm DR}$ must cancel in the final result.

We write therefore
\begin{align}
I_\alpha = M_{\rm DR}^{2\alpha}&\int_0^{\infty} \rd \tau  \left[ \frac{e^{-\mu^2\tau}}{\tau^{(1-\alpha)}} - \frac{2}{\tL^4} \frac{e^{- m^2\tau}}{\tau^{3-\alpha}}
+\frac{2}{\tL^4} \frac{e^{-(\tL^2+m^2)\tau}}{\tau^{3-\alpha}} +\frac{2}{\tL^2} \frac{e^{-(\tL^2+m^2)\tau}}{\tau^{2-\alpha}}  \right]\nonumber\\
= M_{\rm DR}^{2\alpha}&\left[ \Gamma(\alpha)\mu^{-2\alpha} - \frac{2}{\tL^4} \Gamma(\alpha-2)m^{2(2-\alpha)}
+ \frac{2}{\tL^4} \Gamma(\alpha-2)(\tL^2+m^2)^{2-\alpha}\right.\nonumber\\
&+ \left.\frac{2}{\tL^2} \Gamma(\alpha-1)(\tL^2+m^2)^{1-\alpha}
\right]
\end{align}
where we used that
\be
\int_0^{\infty}  \frac{e^{-x \tau}}{\tau^{n-\alpha}} d \tau=x^{n-\alpha-1} \Gamma(1-n+\alpha).
\ee
We can expand this around $\alpha=0$ using that
\be
  M_{\rm DR}^{2\alpha}\Gamma(\alpha-n) x^{n-\alpha} = \frac{(-1)^{n+1} x^n}{n! } \left[-\frac1{\alpha} + \ln\left(\frac{x e^\gamma}{M_{\rm DR}^2}\right) -H_n + O(\alpha)\right],
\ee
where $H_0=0$, $H_n=\sum_{k=1}^n 1/k$ are the harmonic numbers.
One can then check that the residue of the pole vanish because
\begin{align}
\frac{2(m^2+\tL^2)}{\tL^2}
= 1-\frac{m^4}{\tL^4} +\frac{(m^2+\tL^2)^2}{\tL^4}.
\end{align}
 This identity shows that the pole at $\alpha=0$ cancels. The finite limit of $I_\alpha$ is also independent of the auxiliary scale $M_{\rm DR}$. For convenience we can choose $M^2_{\rm DR}=e^\gamma(\tL^2+m^2)$.
One gets that
\begin{align}
I_\alpha
&= -
\ln\qty(\frac{\mu^2 e^\gamma}{M^2_{\rm DR}})
+
\frac{m^4}{\tL^4}\left[
\ln\qty(\frac{m^2 e^\gamma}{M^2_{\rm DR}})-\frac32\right]
-
\qty(\frac{\tL^2+m^2}{\tL^2})^2
\left[
\ln\qty(\frac{(\tL^2+m^2) e^\gamma}{M^2_{\rm DR}})-\frac32\right]
\nonumber\\
& +2
\qty(\frac{\tL^2+m^2}{\tL^2})\left[
\ln\qty(\frac{(\tL^2+m^2) e^\gamma}{M^2_{\rm DR}}) - 1\right]
\nonumber\\
& = -\left[
\ln\qty(\frac{\mu^2}{\tL^2+m^2})
-
\frac{m^4}{\tL^4}
\ln\qty(\frac{m^2 }{\tL^2+m^2})
 + \frac12 -\frac{m^2}{\tL^2}
\right].
\end{align}
 which reproduces \eqref{Gammaex}, since $\Gamma_1=\frac{\mathcal N}{2}\lim_{\alpha\to0}I_\alpha$ and $\mu^2=m^2+\varepsilon^2$.

Thus the analytically continued proper-time integral reproduces exactly the result obtained above by evaluating the finite expression directly. The poles at $\alpha=0$ as well as the dependence on the auxiliary scale $M_{\rm DR} $ cancel after all contributions are combined. This provides a useful consistency check of the calculation and shows that the final result does not depend on the particular intermediate prescription used to treat the singular terms separately.

\bibliographystyle{uiuchept}
\bibliography{biblio}
\end{document}

%% file: macro.tex
\newtheoremstyle{spaced}
{.8em}   
{.2em}   
{\itshape} 
{}        
{\bfseries} 
{.}       
{.5em}    
{}        

\theoremstyle{spaced}

\newcommand{\be}{
\begin{equation}}
  \newcommand{\ee}{
\end{equation}}
\newcommand{\baa}{
\begin{align}}
  \newcommand{\eaa}{
\end{align}}
\newcommand{\bea}{
\begin{eqnarray}}
  \newcommand{\eea}{
\end{eqnarray}}
\newcommand{\pa}{\partial}
\newcommand{\cN}{\mathcal{N}}
\newcommand{\cE}{\mathcal{E}}

\newcommand{\rd}{\mathrm{d}}

\newcommand{\tL}{\tilde{L}}

\makeatletter
\newcommand{\normord@delim}{%
  \mspace{1mu}%
  \vphantom{|}
  \smash{%
    \vcenter{%
      \offinterlineskip
      \hbox{\scalebox{0.55}{$\scriptscriptstyle\bullet$}}%
      \vskip .7ex
      \hbox{\scalebox{0.55}{$\scriptscriptstyle\bullet$}}%
    }%
  }%
  \mspace{1mu}%
}

\newcommand{\normord}[1]{%
  \vphantom{#1}\mathord{%
    \vphantom{|}
    \smash{%
      \mathinner{%
        \mathopen{\normord@delim}%
        \mkern1mu#1\mkern1mu%
        \mathclose{\normord@delim}%
      }%
    }%
  }%
}
\newcommand{\revnormord@delim}{%
  \mspace{1mu}%
  \vphantom{|}%
  \smash{%
    \vcenter{%
      \offinterlineskip
      \hbox{\scalebox{0.8}{$\scriptscriptstyle\boldsymbol{\times}$}}%
      \vskip .2ex
      \hbox{\scalebox{0.8}{$\scriptscriptstyle\boldsymbol{\times}$}}%
    }%
  }%
  \mspace{1mu}%
}

\newcommand{\revnormord}[1]{%
  \vphantom{#1}\mathord{%
    \vphantom{|}%
    \smash{%
      \mathinner{%
        \mathopen{\revnormord@delim}%
        \mkern0.6mu#1\mkern0.6mu%
        \mathclose{\revnormord@delim}%
      }%
    }%
  }%
}

\newcommand{\partcovnormord@delim}{%
  \mspace{1.5mu}%
  \vphantom{|}%
  \smash{%
    \vcenter{%
      \offinterlineskip
      \hbox{\scalebox{1.1}{$\scriptscriptstyle\star$}}%
      \vskip .35ex
      \hbox{\hspace*{0.21ex}\scalebox{0.55}{$\scriptscriptstyle\bullet$}}%
      \vskip .05ex
    }%
  }%
  \mspace{1.5mu}%
}

\newcommand{\partcovnormord}[1]{%
  \vphantom{#1}\mathord{%
    \vphantom{|}%
    \smash{%
      \mathinner{%
        \mathopen{\partcovnormord@delim}%
        #1%
        \mathclose{\partcovnormord@delim}%
      }%
    }%
  }%
}

\newcommand{\covnormord@delim}{%
  \mspace{1.5mu}%
  \vphantom{|}%
  \smash{%
    \vcenter{%
      \offinterlineskip
      \hbox{\scalebox{1.1}{$\scriptscriptstyle\star$}}%
      \vskip .35ex
      \hbox{\scalebox{1.1}{$\scriptscriptstyle\star$}}%
      \vskip .05ex
    }%
  }%
  \mspace{1.5mu}%
}

\newcommand{\covnormord}[1]{%
  \vphantom{#1}\mathord{%
    \vphantom{|}%
    \smash{%
      \mathinner{%
        \mathopen{\covnormord@delim}%
        #1%
        \mathclose{\covnormord@delim}%
      }%
    }%
  }%
}
\makeatother

\usepackage{expl3,xparse}

\ExplSyntaxOn
\NewDocumentCommand{\Operator}{m}
{
  \tl_if_single:nTF { #1 }
  { \hat{#1} }
  { \widehat{#1} }
}

\NewDocumentCommand{\OperatorGaugeFixed}{m}
{
  \tl_if_single:nTF { #1 }
  { \hat{\bm{#1}} }
  { \widehat{\bm{#1}} }
}
\ExplSyntaxOff

\makeatletter
\newcommand\footnoteref[1]{\protected@xdef\@thefnmark{\ref{#1}}\@footnotemark}
\makeatother

\titleformat{\paragraph}
{\normalfont\normalsize\bfseries}{\theparagraph}{1em}{}
\titlespacing*{\paragraph}
{0pt}{3.25ex plus 1ex minus .2ex}{1.5ex plus .2ex}

\newcommand{\stl}{\tilde{\mathscr{l}}}
\newcommand{\sm}{\mathscr{m}}
\newcommand{\sr}{\mathscr{r}}

\newcommand{\one}{\mathbf 1}

\usepackage[normalem]{ulem}

%% file: biblio.bib
@article{Padilla:2015aaa,
    author = "Padilla, Antonio",
    title = "{Lectures on the Cosmological Constant Problem}",
    eprint = "1502.05296",
    archivePrefix = "arXiv",
    primaryClass = "hep-th",
    month = "2",
    year = "2015"
}

@article{Maldacena:1997re,
    author = "Maldacena, Juan Martin",
    title = "{The Large $N$ limit of superconformal field theories and supergravity}",
    eprint = "hep-th/9711200",
    archivePrefix = "arXiv",
    reportNumber = "HUTP-97-A097, HUTP-98-A097",
    doi = "10.4310/ATMP.1998.v2.n2.a1",
    journal = "Adv. Theor. Math. Phys.",
    volume = "2",
    pages = "231--252",
    year = "1998"
}

@article{Gubser:1998bc,
    author = "Gubser, S. S. and Klebanov, Igor R. and Polyakov, Alexander M.",
    title = "{Gauge theory correlators from noncritical string theory}",
    eprint = "hep-th/9802109",
    archivePrefix = "arXiv",
    reportNumber = "PUPT-1767",
    doi = "10.1016/S0370-2693(98)00377-3",
    journal = "Phys. Lett. B",
    volume = "428",
    pages = "105--114",
    year = "1998"
}

@article{Witten:1998qj,
    author = "Witten, Edward",
    title = "{Anti de Sitter space and holography}",
    eprint = "hep-th/9802150",
    archivePrefix = "arXiv",
    reportNumber = "IASSNS-HEP-98-15",
    doi = "10.4310/ATMP.1998.v2.n2.a2",
    journal = "Adv. Theor. Math. Phys.",
    volume = "2",
    pages = "253--291",
    year = "1998"
}

@article{Weinberg:2021exr,
    author = "Weinberg, Steven",
    title = "{On the Development of Effective Field Theory}",
    eprint = "2101.04241",
    archivePrefix = "arXiv",
    primaryClass = "hep-th",
    reportNumber = "UTTG-25-2020",
    doi = "10.1140/epjh/s13129-021-00004-x",
    journal = "Eur. Phys. J. H",
    volume = "46",
    number = "1",
    pages = "6",
    year = "2021"
}

@article{tHooft:1993dmi,
    author = "'t Hooft, Gerard",
    title = "{Dimensional reduction in quantum gravity}",
    eprint = "gr-qc/9310026",
    archivePrefix = "arXiv",
    reportNumber = "THU-93-26",
    journal = "Conf. Proc. C",
    volume = "930308",
    pages = "284--296",
    year = "1993"
}

@article{Susskind:1994vu,
    author = "Susskind, Leonard",
    title = "{The World as a hologram}",
    eprint = "hep-th/9409089",
    archivePrefix = "arXiv",
    reportNumber = "SU-ITP-94-33",
    doi = "10.1063/1.531249",
    journal = "J. Math. Phys.",
    volume = "36",
    pages = "6377--6396",
    year = "1995"
}

@article{Weinberg:1988cp,
    author = "Weinberg, Steven",
    editor = "Hsu, Jong-Ping and Fine, D.",
    title = "{The Cosmological Constant Problem}",
    reportNumber = "UTTG-12-88",
    doi = "10.1103/RevModPhys.61.1",
    journal = "Rev. Mod. Phys.",
    volume = "61",
    pages = "1--23",
    year = "1989"
}

@article{Weinberg:1978kz,
    author = "Weinberg, Steven",
    editor = "Deser, S.",
    title = "{Phenomenological Lagrangians}",
    reportNumber = "HUTP-78-A051A",
    doi = "10.1016/0378-4371(79)90223-1",
    journal = "Physica A",
    volume = "96",
    number = "1-2",
    pages = "327--340",
    year = "1979"
}

@article{Wilson:1973jj,
    author = "Wilson, K. G. and Kogut, John B.",
    title = "{The Renormalization group and the epsilon expansion}",
    doi = "10.1016/0370-1573(74)90023-4",
    journal = "Phys. Rept.",
    volume = "12",
    pages = "75--199",
    year = "1974"
}

@article{Wilson:1971bg,
    author = "Wilson, Kenneth G.",
    title = "{Renormalization group and critical phenomena. 1. Renormalization group and the Kadanoff scaling picture}",
    doi = "10.1103/PhysRevB.4.3174",
    journal = "Phys. Rev. B",
    volume = "4",
    pages = "3174--3183",
    year = "1971"
}

@article{Wilson:1971dh,
    author = "Wilson, Kenneth G.",
    title = "{Renormalization group and critical phenomena. 2. Phase space cell analysis of critical behavior}",
    doi = "10.1103/PhysRevB.4.3184",
    journal = "Phys. Rev. B",
    volume = "4",
    pages = "3184--3205",
    year = "1971"
}

@article{Gibbons:1977mu,
    author = "Gibbons, G. W. and Hawking, S. W.",
    title = "{Cosmological Event Horizons, Thermodynamics, and Particle Creation}",
    doi = "10.1103/PhysRevD.15.2738",
    journal = "Phys. Rev. D",
    volume = "15",
    pages = "2738--2751",
    year = "1977"
}

@book{Gibbons:1994cg,
    editor = "Gibbons, G. W. and Hawking, S. W.",
    title = "{Euclidean quantum gravity}",
    year = "1994"
}

@article{Polchinski:1983gv,
    author = "Polchinski, Joseph",
    title = "{Renormalization and Effective Lagrangians}",
    reportNumber = "HUTP-83-A018",
    doi = "10.1016/0550-3213(84)90287-6",
    journal = "Nucl. Phys. B",
    volume = "231",
    pages = "269--295",
    year = "1984"
}

@inproceedings{Weinberg:2000yb,
    author = "Weinberg, Steven",
    title = "{The Cosmological constant problems}",
    booktitle = "{4th International Symposium on Sources and Detection of Dark Matter in the Universe (DM 2000)}",
    eprint = "astro-ph/0005265",
    archivePrefix = "arXiv",
    reportNumber = "UTTG-07-00",
    doi = "10.1007/978-3-662-04587-9_2",
    pages = "18--26",
    month = "2",
    year = "2000"
}

@article{Appelquist:1974tg,
    author = "Appelquist, Thomas and Carazzone, J.",
    title = "{Infrared Singularities and Massive Fields}",
    reportNumber = "Print-74-1486 (HARVARD)",
    doi = "10.1103/PhysRevD.11.2856",
    journal = "Phys. Rev. D",
    volume = "11",
    pages = "2856",
    year = "1975"
}

@article{Donoghue:2009mn,
    author = "Donoghue, John F.",
    editor = "Pich, A. and Portoles, J. and Rodrigo, G.",
    title = "{When Effective Field Theories Fail}",
    eprint = "0909.0021",
    archivePrefix = "arXiv",
    primaryClass = "hep-ph",
    doi = "10.22323/1.069.0001",
    journal = "PoS",
    volume = "EFT09",
    pages = "001",
    year = "2009"
}

@article{Wilson:1983xri,
    author = "Wilson, K. G.",
    title = "{The renormalization group and critical phenomena}",
    doi = "10.1103/RevModPhys.55.583",
    journal = "Rev. Mod. Phys.",
    volume = "55",
    pages = "583--600",
    year = "1983"
}

@article{Wilson:1974mb,
    author = "Wilson, Kenneth G.",
    title = "{The Renormalization Group: Critical Phenomena and the Kondo Problem}",
    reportNumber = "CLNS-296",
    doi = "10.1103/RevModPhys.47.773",
    journal = "Rev. Mod. Phys.",
    volume = "47",
    pages = "773",
    year = "1975"
}

@article{Isidori:2023pyp,
    author = "Isidori, Gino and Wilsch, Felix and Wyler, Daniel",
    title = "{The standard model effective field theory at work}",
    eprint = "2303.16922",
    archivePrefix = "arXiv",
    primaryClass = "hep-ph",
    reportNumber = "ZU-TH 14/23",
    doi = "10.1103/RevModPhys.96.015006",
    journal = "Rev. Mod. Phys.",
    volume = "96",
    number = "1",
    pages = "015006",
    year = "2024"
}

@article{Weinberg:2009bg,
    author = "Weinberg, Steven",
    title = "{Effective Field Theory, Past and Future}",
    eprint = "0908.1964",
    archivePrefix = "arXiv",
    primaryClass = "hep-th",
    reportNumber = "UTTG-09-09, TCC-028-09",
    doi = "10.22323/1.086.0001",
    journal = "PoS",
    volume = "CD09",
    pages = "001",
    year = "2009"
}

@article{Friedan:1980jm,
    author = "Friedan, Daniel Harry",
    title = "{Nonlinear Models in Two + Epsilon Dimensions}",
    reportNumber = "LBL-11517, UMI-81-13038",
    doi = "10.1016/0003-4916(85)90384-7",
    journal = "Annals Phys.",
    volume = "163",
    pages = "318",
    year = "1985"
}

@article{deBoer:1999tgo,
    author = "de Boer, Jan and Verlinde, Erik P. and Verlinde, Herman L.",
    title = "{On the holographic renormalization group}",
    eprint = "hep-th/9912012",
    archivePrefix = "arXiv",
    reportNumber = "PUPT-1898, ITFA-99-39, SPIN-1999-29",
    doi = "10.1088/1126-6708/2000/08/003",
    journal = "JHEP",
    volume = "08",
    pages = "003",
    year = "2000"
}

@article{Strominger:2001gp,
    author = "Strominger, Andrew",
    title = "{Inflation and the dS / CFT correspondence}",
    eprint = "hep-th/0110087",
    archivePrefix = "arXiv",
    doi = "10.1088/1126-6708/2001/11/049",
    journal = "JHEP",
    volume = "11",
    pages = "049",
    year = "2001"
}

@article{Balasubramanian:2001nb,
    author = "Balasubramanian, Vijay and de Boer, Jan and Minic, Djordje",
    title = "{Mass, entropy and holography in asymptotically de Sitter spaces}",
    eprint = "hep-th/0110108",
    archivePrefix = "arXiv",
    reportNumber = "VPI-IPPAP-01-01, UPR-964-T",
    doi = "10.1103/PhysRevD.65.123508",
    journal = "Phys. Rev. D",
    volume = "65",
    pages = "123508",
    year = "2002"
}

@article{Coleman:1969sm,
    author = "Coleman, Sidney R. and Wess, J. and Zumino, Bruno",
    title = "{Structure of phenomenological Lagrangians. 1.}",
    doi = "10.1103/PhysRev.177.2239",
    journal = "Phys. Rev.",
    volume = "177",
    pages = "2239--2247",
    year = "1969"
}

@article{Callan:1969sn,
    author = "Callan, Jr., Curtis G. and Coleman, Sidney R. and Wess, J. and Zumino, Bruno",
    title = "{Structure of phenomenological Lagrangians. 2.}",
    doi = "10.1103/PhysRev.177.2247",
    journal = "Phys. Rev.",
    volume = "177",
    pages = "2247--2250",
    year = "1969"
}

@article{Chamseddine:2012sw,
    author = "Chamseddine, Ali H. and Connes, Alain",
    title = "{Resilience of the Spectral Standard Model}",
    eprint = "1208.1030",
    archivePrefix = "arXiv",
    primaryClass = "hep-ph",
    doi = "10.1007/JHEP09(2012)104",
    journal = "JHEP",
    volume = "09",
    pages = "104",
    year = "2012"
}

@article{Weinberg:1968de,
    author = "Weinberg, Steven",
    title = "{Nonlinear realizations of chiral symmetry}",
    doi = "10.1103/PhysRev.166.1568",
    journal = "Phys. Rev.",
    volume = "166",
    pages = "1568--1577",
    year = "1968"
}

@article{Peskin:2025lsg,
    author = "Peskin, Michael E.",
    title = "{What is the Hierarchy Problem?}",
    eprint = "2505.00694",
    archivePrefix = "arXiv",
    primaryClass = "hep-ph",
    doi = "10.1016/j.nuclphysb.2025.116971",
    journal = "Nucl. Phys. B",
    volume = "1018",
    pages = "116971",
    year = "2025"
}

@article{Craig:2022eqo,
    author = "Craig, Nathaniel",
    title = "{Naturalness: past, present, and future}",
    eprint = "2205.05708",
    archivePrefix = "arXiv",
    primaryClass = "hep-ph",
    doi = "10.1140/epjc/s10052-023-11928-7",
    journal = "Eur. Phys. J. C",
    volume = "83",
    number = "9",
    pages = "825",
    year = "2023"
}

@article{Becker:2020mjl,
    author = "Becker, Maximilian and Reuter, Martin",
    title = "{Background Independent Field Quantization with Sequences of Gravity-Coupled Approximants}",
    eprint = "2008.09430",
    archivePrefix = "arXiv",
    primaryClass = "gr-qc",
    doi = "10.1103/PhysRevD.102.125001",
    journal = "Phys. Rev. D",
    volume = "102",
    number = "12",
    pages = "125001",
    year = "2020"
}

@article{Becker:2021pwo,
    author = "Becker, Maximilian and Reuter, Martin",
    title = "{Background independent field quantization with sequences of gravity-coupled approximants. II. Metric fluctuations}",
    eprint = "2109.09496",
    archivePrefix = "arXiv",
    primaryClass = "hep-th",
    doi = "10.1103/PhysRevD.104.125008",
    journal = "Phys. Rev. D",
    volume = "104",
    number = "12",
    pages = "125008",
    year = "2021"
}

@article{Ferrero:2025ugd,
    author = "Ferrero, Renata and Naso, Vincenzo and Percacci, Roberto",
    title = "{Quantum Fields and the Cosmological Constant}",
    eprint = "2503.17203",
    archivePrefix = "arXiv",
    primaryClass = "hep-th",
    doi = "10.3390/universe11060173",
    journal = "Universe",
    volume = "11",
    number = "6",
    pages = "173",
    year = "2025"
}

@article{Freidel:2022ryr,
    author = "Freidel, Laurent and Kowalski-Glikman, Jerzy and Leigh, Robert G. and Minic, Djordje",
    title = "{Vacuum energy density and gravitational entropy}",
    eprint = "2212.00901",
    archivePrefix = "arXiv",
    primaryClass = "hep-th",
    doi = "10.1103/PhysRevD.107.126016",
    journal = "Phys. Rev. D",
    volume = "107",
    number = "12",
    pages = "126016",
    year = "2023"
}

@article{Freidel:2023ytq,
    author = "Freidel, Laurent and Kowalski-Glikman, Jerzy and Leigh, Robert G. and Minic, Djordje",
    title = "{On the inevitable lightness of vacuum}",
    eprint = "2303.17495",
    archivePrefix = "arXiv",
    primaryClass = "hep-th",
    doi = "10.1142/S021827182342004X",
    journal = "Int. J. Mod. Phys. D",
    volume = "32",
    number = "14",
    pages = "2342004",
    year = "2023"
}

@article{Freidel:2016pls,
    author = "Freidel, Laurent and Leigh, Robert G. and Minic, Djordje",
    title = "{Quantum Spaces are Modular}",
    eprint = "1606.01829",
    archivePrefix = "arXiv",
    primaryClass = "hep-th",
    doi = "10.1103/PhysRevD.94.104052",
    journal = "Phys. Rev. D",
    volume = "94",
    number = "10",
    pages = "104052",
    year = "2016"
}

@inproceedings{Berglund:2022qcc,
    author = "Berglund, Per and Freidel, Laurent and Hubsch, Tristan and Kowalski-Glikman, Jerzy and Leigh, Robert G. and Mattingly, David and Minic, Djordje",
    title = "{Infrared Properties of Quantum Gravity: UV/IR Mixing, Gravitizing the Quantum -- Theory and Observation}",
    booktitle = "{Snowmass 2021}",
    eprint = "2202.06890",
    archivePrefix = "arXiv",
    primaryClass = "hep-th",
    month = "2",
    year = "2022"
}

@article{Gell-Mann:1954yli,
    author = "Gell-Mann, Murray and Low, F. E.",
    title = "{Quantum electrodynamics at small distances}",
    doi = "10.1103/PhysRev.95.1300",
    journal = "Phys. Rev.",
    volume = "95",
    pages = "1300--1312",
    year = "1954"
}

@article{Stueckelberg:1952hma,
    author = "Stueckelberg, Ernst Carl Gerlach and Petermann, Andreas",
    title = "{Normalization of constants in the quanta theory}",
    doi = "10.5169/seals-112426",
    journal = "Helv. Phys. Acta",
    volume = "26",
    pages = "499--520",
    year = "1953"
}

@article{Callan:1970yg,
    author = "Callan, Jr., Curtis G.",
    title = "{Broken scale invariance in scalar field theory}",
    doi = "10.1103/PhysRevD.2.1541",
    journal = "Phys. Rev. D",
    volume = "2",
    pages = "1541--1547",
    year = "1970"
}

@book{Bogolyubov:1959bfo,
    author = "Bogolyubov, N. N. and Shirkov, D. V.",
    title = "{Introduction to the Theory of Quantized Fields}",
    volume = "3",
    year = "1959"
}

@book{Weinberg:1995mt,
    author = "Weinberg, Steven",
    title = "{The Quantum Theory of Fields. Vol. 1: Foundations}",
    doi = "10.1017/CBO9781139644167",
    isbn = "978-0-521-67053-1, 978-0-511-25204-4",
    publisher = "Cambridge University Press",
    month = "6",
    year = "2005"
}

@book{Weinberg:1996kr,
    author = "Weinberg, Steven",
    title = "{The Quantum Theory of Fields. Vol. 2: Modern applications}",
    doi = "10.1017/CBO9781139644174",
    isbn = "978-1-139-63247-8, 978-0-521-67054-8, 978-0-521-55002-4",
    publisher = "Cambridge University Press",
    month = "8",
    year = "2013"
}

@article{Symanzik:1970rt,
    author = "Symanzik, K.",
    title = "{Small distance behavior in field theory and power counting}",
    reportNumber = "DESY-70-20",
    doi = "10.1007/BF01649434",
    journal = "Commun. Math. Phys.",
    volume = "18",
    pages = "227--246",
    year = "1970"
}

@book{Schwartz:2014sze,
    author = "Schwartz, Matthew D.",
    title = "{Quantum Field Theory and the Standard Model}",
    doi = "10.1017/9781139540940",
    isbn = "978-1-107-03473-0, 978-1-107-03473-0",
    publisher = "Cambridge University Press",
    month = "3",
    year = "2014"
}

@article{Curtright:2011vw,
    author = "Curtright, Thomas L. and Zachos, Cosmas K.",
    title = "{Quantum Mechanics in Phase Space}",
    eprint = "1104.5269",
    archivePrefix = "arXiv",
    primaryClass = "physics.hist-ph",
    reportNumber = "ANL-HEP-PR-11-31, UMTG-22, ANL-HEP-PR-11-31 and UMTG-22",
    doi = "10.1142/S2251158X12000069",
    journal = "Asia Pac. Phys. Newslett.",
    volume = "1",
    pages = "37--46",
    year = "2012"
}

@article{Susskind:1998dq,
    author = "Susskind, Leonard and Witten, Edward",
    title = "{The Holographic bound in anti-de Sitter space}",
    eprint = "hep-th/9805114",
    archivePrefix = "arXiv",
    reportNumber = "SU-ITP-98-39, IASSNS-HEP-98-44",
    month = "5",
    year = "1998"
}

@inproceedings{Hubsch:2026lot,
    author = {H{\"u}bsch, Tristan and Minic, Djordje},
    title = "{Quantum Spacetime, Quantum Gravity and Gravitized Quantum Theory}",
    eprint = "2604.19418",
    archivePrefix = "arXiv",
    primaryClass = "gr-qc",
    month = "4",
    year = "2026"
}

@article{hubsch2024quantum,
  title={Quantum gravity as gravitized quantum theory},
  author={H{\"u}bsch, Tristan and Minic, Djordje},
  journal={arXiv preprint arXiv:2407.06207},
  year={2024}
}

@article{Freidel:2015pka,
    author = "Freidel, Laurent and Leigh, Robert G. and Minic, Djordje",
    title = "{Metastring Theory and Modular Space-time}",
    eprint = "1502.08005",
    archivePrefix = "arXiv",
    primaryClass = "hep-th",
    doi = "10.1007/JHEP06(2015)006",
    journal = "JHEP",
    volume = "06",
    pages = "006",
    year = "2015"
}

@article{Freidel:2018apz,
    author = "Freidel, Laurent and Kowalski-Glikman, Jerzy and Leigh, Robert G. and Minic, Djordje",
    title = "{Theory of metaparticles}",
    eprint = "1812.10821",
    archivePrefix = "arXiv",
    primaryClass = "hep-th",
    doi = "10.1103/PhysRevD.99.066011",
    journal = "Phys. Rev. D",
    volume = "99",
    number = "6",
    pages = "066011",
    year = "2019"
}

@book{aharonov2008quantum,
  title={Quantum paradoxes: quantum theory for the perplexed},
  author={Aharonov, Yakir and Rohrlich, Daniel},
  year={2008},
  publisher={John Wiley \& Sons}
}

@article{Branchina:2025hen,
    author = "Branchina, Carlo and Branchina, Vincenzo and Contino, Filippo and Gandolfo, Riccardo and Pernace, Arcangelo",
    title = "{Gravity and the Higgs boson mass}",
    eprint = "2507.13832",
    archivePrefix = "arXiv",
    primaryClass = "hep-th",
    month = "7",
    year = "2025"
}

@article{fradkin1973s,
  title={S Matrix for Gravitational Field. II. Local measure; General relations; Elements of renormalization theory},
  author={Fradkin, ES and Vilkovisky, GA},
  journal={Physical Review D},
  volume={8},
  number={12},
  pages={4241},
  year={1973},
  publisher={APS}
}

@article{Gibbons:1976ue,
    author = "Gibbons, G. W. and Hawking, S. W.",
    title = "{Action Integrals and Partition Functions in Quantum Gravity}",
    reportNumber = "PRINT-76-0995 (CAMBRIDGE)",
    doi = "10.1103/PhysRevD.15.2752",
    journal = "Phys. Rev. D",
    volume = "15",
    pages = "2752--2756",
    year = "1977"
}

@article{Planck:2018vyg,
    author = "Aghanim, N. and others",
    collaboration = "Planck",
    title = "{Planck 2018 results. VI. Cosmological parameters}",
    eprint = "1807.06209",
    archivePrefix = "arXiv",
    primaryClass = "astro-ph.CO",
    doi = "10.1051/0004-6361/201833910",
    journal = "Astron. Astrophys.",
    volume = "641",
    pages = "A6",
    year = "2020",
    note = "[Erratum: Astron.Astrophys. 652, C4 (2021)]"
}

@article{Branchina:2026vvk,
    author = "Branchina, V. and Contino, F. and Gandolfo, R. and Pernace, A.",
    title = "{Hard cutoff and gauge theories}",
    eprint = "2605.25827",
    archivePrefix = "arXiv",
    primaryClass = "hep-th",
    month = "5",
    year = "2026"
}

@article{Branchina:2024xzh,
    author = "Branchina, Carlo and Branchina, Vincenzo and Contino, Filippo and Pernace, Arcangelo",
    title = "{Path integral measure and the cosmological constant}",
    eprint = "2412.10194",
    archivePrefix = "arXiv",
    primaryClass = "hep-th",
    doi = "10.1103/PhysRevD.111.105018",
    journal = "Phys. Rev. D",
    volume = "111",
    number = "10",
    pages = "105018",
    year = "2025"
}

@article{Branchina:2024lai,
    author = "Branchina, Carlo and Branchina, Vincenzo and Contino, Filippo and Pernace, Arcangelo",
    title = "{Path integral measure and RG equations for gravity}",
    eprint = "2412.14108",
    archivePrefix = "arXiv",
    primaryClass = "hep-th",
    doi = "10.1103/wqv2-j5dt",
    journal = "Phys. Rev. D",
    volume = "111",
    number = "12",
    pages = "125021",
    year = "2025"
}

@article{Ferrero:2024yvw,
    author = "Ferrero, Renata and Percacci, Roberto",
    title = "{The cosmological constant problem and the effective potential of a gravity-coupled scalar}",
    eprint = "2404.12357",
    archivePrefix = "arXiv",
    primaryClass = "hep-th",
    doi = "10.1007/JHEP09(2024)074",
    journal = "JHEP",
    volume = "09",
    pages = "074",
    year = "2024"
}

@article{Cadoni:2026qss,
    author = "Cadoni, Mariano and Herres, Lorenzo and Orlando, Lorenzo and Pitzalis, Mirko",
    title = "{Microscopic entropy of de Sitter spacetime and entropic solution to the old cosmological constant problem}",
    eprint = "2606.23522",
    archivePrefix = "arXiv",
    primaryClass = "hep-th",
    month = "6",
    year = "2026"
}

@article{Ivrii:2016,
    author = {Ivrii, Victor},
    title = {100 years of Weyl's law},
    journal = {Bull. Math. Sci.},
    volume = {6},
    number = {3},
    pages = {379--452},
    year = {2016},
    eprint = {1608.03963},
    archivePrefix = {arXiv},
    primaryClass = {math.SP}
}

@article{Vassilevich:2003xt,
    author = "Vassilevich, D. V.",
    title = "{Heat kernel expansion: User's manual}",
    eprint = "hep-th/0306138",
    archivePrefix = "arXiv",
    doi = "10.1016/j.physrep.2003.09.002",
    journal = "Phys. Rept.",
    volume = "388",
    pages = "279--360",
    year = "2003"
}

@article{Branchina:2025lqw,
    author = "Branchina, Carlo and Branchina, Vincenzo and Contino, Filippo and Gandolfo, Riccardo and Pernace, Arcangelo",
    title = "{Diffeomorphism invariance of the effective gravitational action}",
    eprint = "2506.05100",
    archivePrefix = "arXiv",
    primaryClass = "hep-th",
    doi = "10.1103/p2cz-qwnm",
    journal = "Phys. Rev. D",
    volume = "112",
    number = "4",
    pages = "045002",
    year = "2025"
}

@article{oleszczuk1994symmetry,
  title={A symmetry-preserving cut-off regularization},
  author={Oleszczuk, Michael},
  journal={Zeitschrift f{\"u}r Physik C Particles and Fields},
  volume={64},
  number={3},
  pages={533--538},
  year={1994},
  publisher={Springer}
}

@article{Bilal:2013iva,
    author = "Bilal, Adel and Ferrari, Frank",
    title = "{Multi-Loop Zeta Function Regularization and Spectral Cutoff in Curved Spacetime}",
    eprint = "1307.1689",
    archivePrefix = "arXiv",
    primaryClass = "hep-th",
    doi = "10.1016/j.nuclphysb.2013.10.003",
    journal = "Nucl. Phys. B",
    volume = "877",
    pages = "956--1027",
    year = "2013"
}

@article{Avramidi:1997jy,
    author = "Avramidi, I. G.",
    title = "{Covariant techniques for computation of the heat kernel}",
    eprint = "hep-th/9704166",
    archivePrefix = "arXiv",
    doi = "10.1142/S0129055X99000295",
    journal = "Rev. Math. Phys.",
    volume = "11",
    pages = "947--980",
    year = "1999"
}

@inproceedings{derezinski2020pseudodifferential,
  title={Pseudodifferential Weyl Calculus on (Pseudo-) Riemannian Manifolds},
  author={Derezi{\'n}ski, Jan and Latosi{\'n}ski, Adam and Siemssen, Daniel},
  booktitle={Annales Henri Poincar{\'e}},
  volume={21},
  number={5},
  pages={1595--1635},
  year={2020},
  organization={Springer}
}

@article{Chamseddine:2009drp,
    author = "Chamseddine, Ali H. and Connes, Alain",
    title = "{The Uncanny Precision of the Spectral Action}",
    eprint = "0812.0165",
    archivePrefix = "arXiv",
    primaryClass = "hep-th",
    doi = "10.1007/s00220-009-0949-3",
    journal = "Commun. Math. Phys.",
    volume = "293",
    pages = "867--897",
    year = "2010"
}

@article{Chamseddine:2010ud,
    author = "Chamseddine, Ali H. and Connes, Alain",
    title = "{Noncommutative Geometry as a Framework for Unification of all Fundamental Interactions including Gravity. Part I}",
    eprint = "1004.0464",
    archivePrefix = "arXiv",
    primaryClass = "hep-th",
    doi = "10.1002/prop.201000069",
    journal = "Fortsch. Phys.",
    volume = "58",
    pages = "553--600",
    year = "2010"
}

@article{andersson2025pseudodifferential,
  title={Pseudodifferential Weyl calculus on vector bundles},
  author={Andersson, Lars and Moser, Benjamin and Oancea, Marius A and Paganini, Claudio F and Schmid, Gabriel},
  journal={arXiv preprint arXiv:2507.11965},
  year={2025}
}

@article{Bousso:2002ju,
    author = "Bousso, Raphael",
    title = "{The Holographic principle}",
    eprint = "hep-th/0203101",
    archivePrefix = "arXiv",
    reportNumber = "NSF-ITP-02-17, NSF-ITP-02-17",
    doi = "10.1103/RevModPhys.74.825",
    journal = "Rev. Mod. Phys.",
    volume = "74",
    pages = "825--874",
    year = "2002"
}

@article{Bousso:2000md,
    author = "Bousso, Raphael",
    title = "{Bekenstein bounds in de Sitter and flat space}",
    eprint = "hep-th/0012052",
    archivePrefix = "arXiv",
    reportNumber = "NSF-ITP-00-123",
    doi = "10.1088/1126-6708/2001/04/035",
    journal = "JHEP",
    volume = "04",
    pages = "035",
    year = "2001"
}

@article{bekenstein1974generalized,
  title={Generalized second law of thermodynamics in black-hole physics},
  author={Bekenstein, Jacob D},
  journal={Physical Review D},
  volume={9},
  number={12},
  pages={3292},
  year={1974},
  publisher={APS}
}

@article{wall2009ten,
  title={Ten proofs of the generalized second law},
  author={Wall, Aron C},
  journal={Journal of High Energy Physics},
  volume={2009},
  number={06},
  pages={021--021},
  year={2009}
}

@article{wall2012proof,
  title={Proof of the generalized second law for rapidly changing fields and arbitrary horizon slices},
  author={Wall, Aron C},
  journal={Physical Review D—Particles, Fields, Gravitation, and Cosmology},
  volume={85},
  number={10},
  pages={104049},
  year={2012},
  publisher={APS}
}
